\documentclass[11pt,headings=big,numbers=noenddot,DIV=14,a4paper]{article}%
\pdfoutput=1

\usepackage[english]{babel}
\usepackage[nosort,compress]{cite}
\usepackage{graphicx,epsfig,subfigure}
\usepackage[dvipsnames]{xcolor}
\usepackage[linktoc=page,bookmarks=false,colorlinks=false,
linkbordercolor=RoyalBlue,citebordercolor=ForestGreen,urlbordercolor=CornflowerBlue]{hyperref}
\usepackage[normal,font=small,labelfont=bf,labelsep=period]{caption}
\usepackage{amsmath,amssymb,amsfonts,array,enumerate,slashed,multirow,fancybox,marvosym,wasysym,soul}
\usepackage[normalem]{ulem}

\usepackage[margin=1in]{geometry}

\font\tenrsfs=rsfs10 at 12pt

\font\sevenrsfs=rsfs7
\font\fiversfs=rsfs5
\newfam\rsfsfam
\textfont\rsfsfam=\tenrsfs
\scriptfont\rsfsfam=\sevenrsfs
\scriptscriptfont\rsfsfam=\fiversfs
\def\mathscr#1{{\fam\rsfsfam\relax#1}}

\font\teneusm=eusm10 at 12pt
\newfam\eusmfam
\textfont\eusmfam=\teneusm

\font\tenbbm=bbm10 at 12pt
\newfam\bbmfam
\textfont\bbmfam=\tenbbm

\def\L{\mathcal{L}}

\def\O{\mathcal{O}}

\def\W{\mathcal{W}}

\def\d{{\rm d}}

\numberwithin{equation}{section}
\newcommand{\sectionname}{section}
\newcommand{\figuresnames}{figure}

\definecolor{darkblue}{rgb}{0,0.2,0.6}
\definecolor{viola}{rgb}{.5,0,.5}
\definecolor{verde}{rgb}{0,.45,0}

\title{\vspace{-2cm}\begin{flushright}
{\small EFI-15-17\hfill FLAVOUR(267104)-ERC-97}
\end{flushright}
\vspace{1cm}
\bf\LARGE Singlet-like Higgs bosons\\ at present and future colliders}
\author{\large Dario Buttazzo$^a$, Filippo Sala$^b$, Andrea Tesi$^c$}
\date{\small {\it  $^a$Institute for Advanced Study \& Physik Department, TUM, 85748 Garching, Germany\\
$^b$Institut de Physique Th\'eorique, Universit\'e Paris Saclay, CNRS, CEA, F-91191 Gif-sur-Yvette, France\\
$^c$Department of Physics, Enrico Fermi Institute, University of Chicago, Chicago, IL 60637}}

\begin{document}

\begin{titlepage}
\maketitle

\thispagestyle{empty}

\begin{abstract}
\noindent
The presence of extra scalar singlets is a feature of several motivated extensions of the Standard Model, and the mixing of such a singlet with the Higgs boson is allowed to be quite large by current experiments. In this paper we perform a thorough phenomenological study of this possibility. We consider both direct and indirect searches, and we quantify the current constraints as well as the prospects for future hadron and lepton machines -- from the forthcoming LHC run up to a futuristic 100 TeV proton-proton collider.
The direct reaches are obtained extrapolating the current limits with a technique that we discuss and check with various tests.
We find a strong complementarity between direct and indirect searches, with the former dominating for lower values of the singlet mass. We also find that the trilinear Higgs coupling can have sizeable deviations from its Standard Model value, a fact for which we provide an analytical understanding. The results are first presented in a general scalar singlet extension of the Standard Model, taking advantage of the very small number of parameters relevant for the phenomenology. Finally, we specify the same analysis to a few most natural models, {\it i.e.}\ the Next-to-Minimal Supersymmetric Standard Model, Twin Higgs and Composite Higgs.
\end{abstract}

\vfill
\noindent\line(1,0){188}\\\medskip
\footnotesize{E-mail: \scriptsize\tt{\href{mailto:dario.buttazzo@cern.ch}{dario.buttazzo@cern.ch}, \href{mailto:filippo.sala@cea.fr}{filippo.sala@cea.fr}, \href{mailto:atesi@uchicago.edu}{atesi@uchicago.edu}}}

\end{titlepage}

\tableofcontents


\section{Introduction}\label{sec:intro}

Is the Higgs boson \cite{Aad:2012tfa,Chatrchyan:2012ufa} the only light scalar particle, or is it rather the low-hanging fruit of a richer scalar sector of Nature?
This question is of paramount importance for the field of particle physics. Discovering another scalar close-by would in fact have deep implications for our understanding of the concept of naturalness\cite{tHooft:1979bh}, challenging frameworks that do not aim at removing the apparent tuning of the weak scale\cite{Agrawal:1997gf}. 

Among the many conceivable scalar extensions of the Standard Model (SM), arguably the simplest one consists in the addition of a singlet under the SM gauge group which mixes with the already discovered Higgs. This possibility is ubiquitous in many new physics scenarios, most notably in two of the most successful proposals to address the hierarchy problem of the Fermi scale: the Next-to-Minimal Supersymmetric Standard Model (NMSSM, see \cite{Ellwanger:2009dp} for a review), and Twin Higgs \cite{Chacko:2005pe}.
It is also interesting that, in the context of ``finite naturalness'' \cite{Farina:2013mla}, an extra singlet-like scalar would also allow to dynamically generate the weak and Dark Matter scales \cite{Hambye:2013dgv}. Moreover, and independently of the hierarchy problem, the same addition to the SM is among the simples possibilities to achieve electroweak baryogengesis \cite{Profumo,Curtin:2014jma,Craig:2014lda}.

From an experimental point of view, the 8 TeV LHC run has given us a knowledge of the main Higgs properties to a 20\% accuracy \cite{Khachatryan:2014jba,ATLAS-CONF-2015-007}, and no direct evidence for new physics whatsoever. Despite that, scenarios exist that are poorly constrained by present data, an extra scalar singlet mixed with the Higgs being a paradigmatic example of this case.

The above considerations strongly motivate a thorough experimental quest for such a particle. The goal of this paper is to provide a general and solid strategy to perform this hunt both at current and future machines.

Several studies in this direction have been performed in the literature (see {\it e.g.}\ the recent \cite{Bertolini:2012gu,Robens:2015gla,Falkowski:2015iwa,Gorbahn:2015gxa}), however they have mostly restricted their attention to some particular direct and/or indirect search, to some particular model, or to some particular machine. In this paper our aim is to be more comprehensive, building a picture that allows to grasp the general phenomenological features, yet in an exhaustive way. In order to do so, we develop an analytical understanding of what the parameters most relevant for the phenomenology are, freeing us from the necessity of numerical scans. In this way we are able to identify the most promising ways to probe this scenario, thus providing key information for current experiments, as well as for planning future ones.

In particular, we will consider several experimental programs besides the second run of the LHC that has just started. Our study will include the high luminosity phase of the LHC itself (HL-LHC \cite{Gianotti:2002xx}), a higher energy phase (HE-LHC), linear (ILC \cite{Baer:2013cma} and CLIC \cite{Linssen:2012hp, Lebrun:2012hj}) as well as circular (FCC-ee \cite{Koratzinos:2013ncw} and CEPC \cite{CEPC}) electron-positron colliders, up to a futuristic 100 TeV proton-proton collider (FCC-hh \cite{100TeV}).

The natural approach to study such an extended Higgs sector relies on two complementary strategies: 
\begin{itemize}
\item[$\circ$] Indirect searches: modification of the couplings of the 125 GeV Higgs boson; 
\item[$\circ$] Direct searches: production and detection of the heavier scalar. 
\end{itemize}
If on one hand the impact of an improvement in precision on the Higgs signal strengths is quite easy to estimate, on the other hand the study of the reach of direct searches requires generally an amount of numerical analysis. In this paper, however, we rely on a technique introduced in \cite{SalamWeiler} and developed in \cite{Thamm:2015zwa} to extrapolate the reach of future colliders from the existing limits of the first LHC run. This will help to get an intuition of the capability of the new experiments, avoiding the use of detailed collider simulations, and it will allow for an easy comparison between the relevance of precision physics and direct searches.

The paper is organised as follows. In \sectionname~\ref{sec:Singlet_generic} we introduce the model and present the basic phenomenological consequences, showing also what are the simplifications that can be made. In \sectionname~\ref{sec:DirectSearches} we review the extrapolation procedure presented in \cite{SalamWeiler, Thamm:2015zwa} discussing its virtues and limitations, and we present our projections for the sensitivity of direct searches in the relevant channels at future machines. These results have a general validity that goes beyond the context of the singlet scenario analysed in this paper. We then apply the bounds to the case of generic singlet scalar coupled to the SM. Sections~\ref{sec:susy} and \ref{sec:twin} are dedicated to the NMSSM, and Twin and Composite Higgs, respectively, and there we project present and future bounds into the relevant parameter space of the two models. We conclude in \sectionname~\ref{sec:conclusions}. Technical details about the extrapolation procedure are given in appendix~\ref{sec:appendix}.


\section{Phenomenology of a scalar singlet}
\label{sec:Singlet_generic}


\subsection{A convenient description}
\label{sec:parametrization}

We add to the SM particle content a scalar field $S$, singlet under the SM gauge group. 
The quantum numbers of the real component of $S$, on which we will focus, allow it to mix with the SM Higgs boson, and we dub the mixing angle $\gamma$.
Let $h$ and $\phi$ be the two neutral CP-even propagating degrees of freedom, $h$ being the one observed at the LHC, and $m_h = 125.1$ GeV \cite{Aad:2015zhl} and $m_\phi$ their physical masses. In terms of the gauge eigenstates
\begin{align}
\label{fields}
H &= \begin{pmatrix}
\pi^+\\
\dfrac{v + h^0 + i\pi^0}{\sqrt{2}}
\end{pmatrix},&
S &= \frac{v_s + s^0}{\sqrt 2},
\end{align}
the mass eigenstates read
\begin{align}\label{rotation}
h &= h^0\cos\gamma + s^0\sin\gamma,&
\phi &= -h^0\sin\gamma + s^0\cos\gamma,
\end{align}
where $v_s$ is the vacuum expectation value (vev) of the singlet field, and $v \simeq 246$ GeV.

All the $h$ and $\phi$ couplings to SM fermions and vectors are fixed, due to the singlet nature of $S$, only by the mixing angle $\gamma$. They can be written in terms of the couplings of a standard Higgs boson of the same mass as
\begin{align}\label{eq:couplings}
\frac{g_{hff}}{g_{hff}^{\rm SM}} &= \frac{g_{hVV}}{g_{hVV}^{\rm SM}} = c_\gamma, & \frac{g_{\phi ff}}{g_{hff}^{\rm SM}} &= \frac{g_{\phi VV}}{g_{hVV}^{\rm SM}} = -s_\gamma\,,
\end{align}
where we introduce the notation $s_\theta \equiv \sin\theta$, $c_\theta \equiv \cos\theta$.
Notice that the branching ratios of $h$ are not modified with respect to their SM values, since all the couplings are rescaled by the same factor; the only observable deviation from the SM is therefore the reduced production cross-section.
Concerning $\phi$, its production cross-section $\sigma_{pp\to \phi}$ is simply the one of a standard Higgs boson of mass $m_{\phi}$ rescaled by $s^2_\gamma$. Its branching ratios are again those of a SM Higgs boson of the same mass below the kinematic threshold $m_\phi = 2 m_h$, where the width $\Gamma_{\phi \to hh}$ can in principle become sizeable. All these considerations can be modified by the presence of operators of dimension greater than four in a strongly coupled theory, but in the following we will consider only weakly coupled scenarios.

The phenomenology of the two scalars can then be summarised as
\begin{align}
\mu_h &= c^2_\gamma \times \mu_{\rm SM},\label{signal_strength_Higgs}\\ 
\mu_{\phi \to VV,ff} &= s^2_\gamma \times \mu_{\rm SM}(m_\phi)\times (1-{\rm BR}_{\phi \to hh})\,,\label{eq:signal_strength}\\
 \mu_{\phi \to hh} &= s^2_\gamma \times \sigma_{\rm SM}(m_\phi)\times {\rm BR}_{\phi \to hh}\,, \label{eq:signal_strength_hh}
\end{align}
where $\mu_{h,\phi}$ are respectively the signal strengths of $h$ and $\phi$ in any SM channel (unless specified), $\mu_{\rm SM}$ is the  signal strength of a SM Higgs boson, and $\sigma_{\rm SM}$ is its production cross-section. The measurement of the $h$ signal strengths, and hence of its couplings, puts a model independent constraint on the mixing angle $\gamma$. This in turn limits the magnitude of possible direct signals of $\phi$, which for large enough $m_\phi$ decays mainly into a pair of vector bosons, $W^+W^-$ and $ZZ$, and into $hh$ if the relative branching ratio is large enough.
The only relevant fermionic decay mode of $\phi$ is a pair of top quarks, which is however subdominant with respect to the previous ones.

One reads in expressions \eqref{eq:signal_strength} and \eqref{eq:signal_strength_hh} that the decays of both scalars are completely determined by just three parameters:
\begin{itemize}
\item[$\diamond$] the mass of the singlet-like scalar $m_\phi$,
\item[$\diamond$] the mixing angle $\gamma$, 
\item[$\diamond$] the branching ratio BR$_{\phi \to hh}$.
\end{itemize}
We now discuss them in order to understand better and simplify the description of the phenomenology, starting with BR$_{\phi \to hh}$.

For $m_\phi \gg m_W$, one knows that decays of $\phi$ into vector bosons are determined by their Goldstone nature, and that they dominate over all the other SM decay modes. This implies the asymptotic relation
\begin{equation}
{\rm BR}_{\phi \to hh} = {\rm BR}_{\phi \to ZZ} = \frac{1}{2}\,{\rm BR}_{\phi \to WW} = \frac{1}{4}\,,\qquad \quad m_\phi \gg m_W.
\label{eq:BRasymptotic}
\end{equation}
For large enough $m_\phi$ the number of free parameters drops then from three to two, further simplifying the description of the model. It is natural to choose $m_\phi$ as one of them.

Concerning the other free parameter, notice that for \textit{any} given scalar potential, the mixing angle $\gamma$ between the two CP-even scalars can be expressed in the form
\begin{equation}\label{singamma}
\sin^2\gamma = \frac{M^2_{hh} - m_{h}^2}{m_{\phi}^2 - m_{h}^2},
\end{equation}
where $M_{hh}^2$ is the (1,1) element of the $2\times2$ mass matrix $M^2$ of the scalar system before diagonalisation, in the $(h^0,s^0)$ basis.
Notice that one has three possibilities to obtain a relation between $\gamma$, the physical masses and the parameters of the Lagrangian: one possibility for each independent entry of the mass matrix $M^2$.
We find it convenient to choose $M_{hh}^2$, because it is always of the order of the electroweak scale $v$, despite its explicit form as a function of the Lagrangian parameters depends on the model under consideration. While this statement will be made more explicit in the discussion of the potential in \sectionname~\ref{sec:potential}, it can be understood qualitatively by noting that, in general, it applies to all cases where the only field responsible for EW symmetry breaking (EWSB) is the one associated with that single entry of $M^2$.\footnote{The only dimensionful parameter of the SM potential is a trivial example of the above property: its size is related to $v$ by an order one factor, the Higgs quartic coupling.} This property holds in particular for the doublet-singlet system under consideration, since the scalar field $S$ does not play any role in EWSB.
We conclude that the use of $M_{hh}^2$ as a free parameter to describe the phenomenology, instead of $\sin^2\gamma$, is a convenient choice.


\subsection{Potential and couplings among the Higgses}\label{sec:potential}

The discussion performed so far allows to treat the decay $\phi \to hh$ only in an approximate manner. Moreover, it has nothing to say about possible deviations of the trilinear Higgs coupling $g_{hhh}$ from its SM value. It is known that  such deviations are allowed to be quite large, at least in some specific models containing singlets\cite{Barbieri:2013hxa}. It would then be interesting to compare them with the other indirect signals, possibly in a most general way.

With this in mind, we write down the most general form of a renormalisable scalar potential involving a doublet $H$ and a singlet $S$ as
\begin{equation}
V(H,S) = \mu_H^2\,|H|^2 + \lambda_H\,|H|^4  + \lambda_{HS}\,S^2 |H|^2 + a_H\,S |H|^2 + \mu_S^2\,S^2 + a_S\,S^3 + \lambda_S\,S^4.\\
\label{potential}
\end{equation}
By deriving the potential with respect to $S$, one sees that the condition $v_s = 0$ for the singlet vev requires $a_H = 0$. This further implies $\sin\gamma = 0$, as one can infer from \eqref{potential}. This is not the case we are interested in, and hence we assume $v_s \neq 0$ in the rest of our discussion. Since $v_s \neq 0$ spontaneously breaks the symmetry $S\to-S$, we assume in general that also $a_H$ and $a_S$ are different from zero. A non zero positive $\lambda_S$ is then a necessary condition for the potential to be bounded from below (neglecting runnings at higher energies that could change this limitation, see {\it e.g.}\ \cite{Hambye:2013dgv}).

We can then substitute a subset of the seven parameters of the potential in \eqref{potential}, with the vevs $v$ and $v_s$ and with the physical masses $m_h$ and $m_\phi$. We start by removing the mass parameters $\mu_H$ and $\mu_S$, in favour of the vevs. In this way one realises that the following relation holds
\begin{equation}
M_{hh}^2 = 2 \lambda_H v^2\,,
\label{Mhh}
\end{equation}
substantiating the discussion of the previous section. We then proceed to exchange the last two dimensionful parameters of \eqref{potential}, $a_H$ and $a_S$, in favour of the physical masses $m_h$ and $m_\phi$. The most general addition of a scalar singlet to the SM can then be described in terms of five free parameters:
\begin{equation}
m_\phi, \qquad M_{hh}, \qquad v_s, \qquad \lambda_{HS}, \qquad \lambda_S\,.
\label{our_parameters}
\end{equation}

The mixing angle $\gamma$, which determines most of the phenomenology of the model, is a function of only two of them, see \eqref{singamma}.
The couplings among the scalars, on the other hand, explicitly depend also on the other three parameters. They read\\
\begin{align}
g_{\phi h h} &= s_\gamma \left[ \frac{\lambda_{HS}\,v}{2}
-\frac{m_\phi^2 + 2 m_h^2}{2v} - \frac{v^2}{v_s^2}\frac{m_\phi^2 - m_h^2}{8 v}
+\frac{s_{2 \gamma}}{2}  \frac{m_\phi^2 + m_h^2 -\lambda_{HS}\,v^2+2 \lambda_S\,v_s^2}{v_s}\right.\notag\\
&\qquad\qquad\,\,\left.- \frac{c_{2 \gamma}}{2} \frac{m_\phi^2 + 2 m_h^2 - 3 \lambda_{HS}\,  v^2}{v}
+\frac{m_\phi^2 - m_h^2}{4 v} \frac{v}{v_s}\left(s_{4\gamma} - \frac{v}{v_s}s_{2\gamma}^2\right)
\right],\label{g211}\\
\frac{g_{hhh}}{g_{hhh}^{\rm SM}} &= c_\gamma  \left[1 + s_\gamma^2 \left(\frac{\lambda_{HS}\,  v^2}{m_h^2} -1\right) - \frac{v^2}{v_s^2} \frac{s_\gamma^4}{3} \left(\frac{m_\phi^2}{m_h^2}-1\right) \right] \notag\\
&\qquad\qquad\!\!+\frac{v}{v_s} \frac{s_\gamma^3}{3} \left[1+ \frac{m_\phi^2-\lambda_{HS}\,  v^2}{m_h^2} + c_{2 \gamma} \left(\frac{m_\phi^2}{m_h^2} -1\right) + 2 \lambda_S \frac{v_s^2}{m_h^2}\right],\label{g111}
\end{align}
where $\gamma$ is as in \eqref{singamma} and $g_{hhh}^{\rm SM}$ is the SM value of the triple Higgs coupling.
One would like to know how $v_s$, $\lambda_{HS}$ and $\lambda_S$ impact the phenomenology, compared to $m_\phi$ and $M_{hh}$: if all the parameters of \eqref{our_parameters} were equally important, as one could expect in principle from \eqref{g211} and \eqref{g111}, the description of this part of the phenomenology would be somehow obscured. We show here that this is not the case, and that the impact of  $v_s$ actually dominates over the one of $\lambda_{HS}$ and $\lambda_S$ in most of the relevant parameter space. We do so by performing a large $m_\phi$ expansion of the interesting phenomenological quantities, BR$_{\phi \to hh}$ and $g_{hhh}/g_{hhh}^{\rm SM}$, which reads
\begin{eqnarray}
{\rm BR}_{\phi \to hh}& = & \frac{1}{4} - \frac{3}{4} \, \frac{v}{v_s}\,\frac{\sqrt{M_{hh}^2-m_h^2}}{m_\phi} \nonumber \\
& &+\, \frac{3}{8} \left( \frac{v^2}{v_s^2}\, \frac{M_{hh}^2 - m_h^2}{m_\phi^2} + \frac{2 m_W^2 +  m_Z^2 + 2 m_h^2 - M_{hh}^2 - 2 \lambda_{HS} v^2}{m_\phi^2}\right)
 + \O\left(\frac{v^3}{m_\phi^3}
\right)\nonumber \\
&= & \frac{1}{4} - \frac{3}{4} \frac{v}{v_s} \sin\gamma + \O\left(\frac{v^2}{m_\phi^2}\right),\label{BRh2hh}\\[5pt]
\frac{g_{hhh}}{g_{hhh}^{\rm SM}} & = & 1 + \frac{2}{3}\frac{v}{v_s} \,\frac{\sqrt{M_{hh}^2-m_h^2}}{m_\phi}\,\left( \frac{M_{hh}^2}{m_h^2} - 1\right) 
\nonumber \\
& & -\, \left(\frac{M_{hh}^2}{m_h^2} - 1\right) \left(\frac{1}{3}\,\frac{v^2}{v_s^2}\, \frac{M_{hh}^2 - m_h^2}{m_\phi^2} + \frac{1}{2}\,\frac{3 m_h^2 - 2 \lambda_{HS} v^2}{m_\phi^2}\right)
+ \O\left(\frac{v^3}{m_\phi^3}\right).
\label{ghhh}
\end{eqnarray}
Here we are assuming that $M_{hh}$ and $v_s$ are of order of the weak scale, and we collectively denote by $\mathcal{O}(v^2/m_\phi^2)$ orders of $\lambda_{HS} v^2 / m_\phi^2$, $\lambda_S v_s^2 / m_\phi^2$, and $M_{hh}^2 / m_{\phi}^2$. We do not keep track of the explicit dependence on the couplings in the higher order terms, since we are considering only weakly coupled scenarios where these couplings are never much greater than one.
A few remarks are in order:
\begin{itemize}
\item[$\circ$] both quantities have the correct asymptotic behaviour if the singlet is decoupled;
\item[$\circ$] at first order in $v/m_\phi$, only $v_s$ determines BR$_{\phi \to hh}$ and $g_{hhh}/g_{hhh}^{\rm SM}$ (besides $m_\phi$ and $M_{hh}$), and at second order only $\lambda_{HS}$ enters the expressions. We have thus obtained an analytical understanding of which extra parameters will impact more on the phenomenology;
\item[$\circ$] if $v/v_s = \O(m_\phi/v)$ or bigger, then both expansions (\ref{BRh2hh}) and (\ref{ghhh}) are inconsistent.

\end{itemize}
In the following, we will always use the complete expressions (\ref{g211}), (\ref{g111}) in our phenomenological studies.
We notice that, choosing a very small value for $v_s/v$, the width $\Gamma_\phi$ of the singlet-like scalar becomes larger than its mass, and the triple Higgs coupling grows as $v^2/v_s^2$, thus making the model effectively strongly coupled. Furthermore, $a_S$ also grows for small $v_s$, introducing a hierarchy among the dimensionful parameters in the potential.

We conclude this section observing that the minimum $(\langle |H| \rangle,\langle S \rangle) = (v,v_s)/\sqrt{2}$ is always the lowest one, for a given choice of the parameters. We verified this in a range that widely covers our phenomenological study.
For a more detailed study of the theoretical constraints on a singlet-Higgs potential, see \cite{Chen:2014ask}.


\subsection{Indirect signals: triple Higgs coupling vs.\ signal strengths}

We have now developed enough tools to study the impact of indirect searches in the scenario under consideration. Before exposing quantitative predictions, it is important to know the present and future experimental situation. We thus begin this section with an overview of the current exclusions, as well as the expectations for future sensitivities in precision Higgs measurements.

\paragraph{Current and future sensitivities to Higgs couplings.}
A global fit to all the Higgs signal strengths as measured at the 8 TeV LHC constrains $\sin^2\gamma\leq 0.23$ at 95\% C.L. \cite{Giardino:2013bma,Falkowski:2013dza}. Concerning future projections, we rely on the comprehensive summary performed at Snowmass in 2013 \cite{Dawson:2013bba}, and we refer the interested reader to it for more details.\footnote{More recent results, when available (see {\it e.g.}\ \cite{ATLAS-CONF-2015-007,ATL-PHYS-PUB-2014-016}), substantially agree with the numbers given here.}

\begin{table}[t]
\renewcommand{\arraystretch}{1.2}
\begin{center}
\small
\begin{tabular}{r|cc}
 1$\sigma$ reach in & $s^2_\gamma$ & $\Big|1-\dfrac{g_{hhh}}{g_{hhh}^{\rm SM}}\Big|$\\ \hline \hline
 LHC8 &  0.2 &  -- \\
 LHC14 &  0.08-0.12 & -- \\
 HL-LHC  &  4-8$\times 10^{-2}$&  0.5 \\
 HE-LHC& -- & 0.2 \\
 FCC-hh &  --& 0.08 \\ \hline
 ILC & 2$\times 10^{-2}$ & 0.21-0.83\\
 ILC-up &  4$\times 10^{-3}$ & 0.13-0.46 \\
 CLIC  & 2-3$\times 10^{-3}$ & 0.1-0.21 \\
 CEPC &  2$\times 10^{-3}$  &  --\\
  FCC-ee &  1$\times 10^{-3}$  &  --
  
\end{tabular}
\caption{\label{tab:Higgs_couplings}\small Current and indicative expected precisions on $s^2_\gamma$, and on the trilinear Higgs coupling $g_{hhh}$. The values for $s_\gamma^2$ correspond to the most precise coupling (always $ZZ$ and/or $WW$). All the numbers are taken from \cite{Dawson:2013bba} (see \cite{CEPC} for CEPC).}
\end{center}
\end{table}
We report in table~\ref{tab:Higgs_couplings} the expected $1\sigma$ precisions in $s^2_\gamma$ coming from the measurement of a single coupling;  for comparison we provide also the current best precision in a single channel from the LHC8 data.
In the case of the mixing with a singlet, these numbers provide at least a good indication of the possible outcome of a global fit. Notice that this is not the case for more complex Higgs sectors, like when the Higgs mixes with another doublet ({\it e.g.}\ MSSM), where a global fit is usually much more constraining (see {\it e.g.}\ \cite{Barbieri:2013hxa}).
The hadronic collider configurations that we consider are LHC at 14 TeV with 300 fb$^{-1}$ (LHC14) and 3000 fb$^{-1}$ (HL-LHC) of integrated luminosity. Assuming one will measure SM central values of the Higgs signal strengths, the expected precision will be in the ballpark of 10\% and 5\% respectively.
The ranges reported in table~\ref{tab:Higgs_couplings} come from different assumptions on the future theory errors \cite{ATL-PHYS-PUB-2014-016}, and from different systematics scenarios \cite{CMS-NOTE-2012-006}. Future projections for higher energy hadronic colliders are currently unavailable, needless to say they would constitute an essential ingredient in the present analysis. For leptonic machines, we report the expected precisions for the ILC and its luminosity upgrade (ILC-up), for CLIC and for circular colliders such as FCC-ee or CEPC. For CLIC we report two numbers, the first assuming a data taking up to 1.4 TeV with 1.5 fb$^{-1}$, the second adding 2 fb$^{-1}$ at 3 TeV.  Assuming SM central values, $s^2_\gamma$ could be constrained from the per-cent down to the per-mille level, FCC-ee being the most sensitive collider in this respect.
Concerning the trilinear Higgs coupling, for definiteness we stick again to the review \cite{Dawson:2013bba}, and refer the interested reader to it for details. The expected $1\sigma$ precisions achievable at future machines are also reported in table~\ref{tab:Higgs_couplings}. The measurement of the trilinear Higgs coupling benefits from higher energy collisions, thus the projection for FCC-ee is not reported. The benchmarks we consider for hadronic ($pp$) colliders are a 33 TeV one (HE-LHC) and a 100 TeV one (FCC-hh), both with 3 ab$^{-1}$ of integrated luminosity. The two numbers given for ILC and ILC-up correspond to the ILC500 and ILC1000 cases, those given for CLIC have the same meaning as for the signal strengths, and assume electron beam polarisation (for unpolarised beams, the worst expected precision would increase to 0.28). Again we warn the reader that such numbers are to be taken with a grain of salt, possibly even more than in the case of signal strengths \cite{Dawson:2013bba}. In any case it looks very challenging to push the sensitivity in $g_{hhh}$ below the $\sim$10\% level.

\begin{center}
\begin{figure}[t]
\includegraphics[width=.48\textwidth]{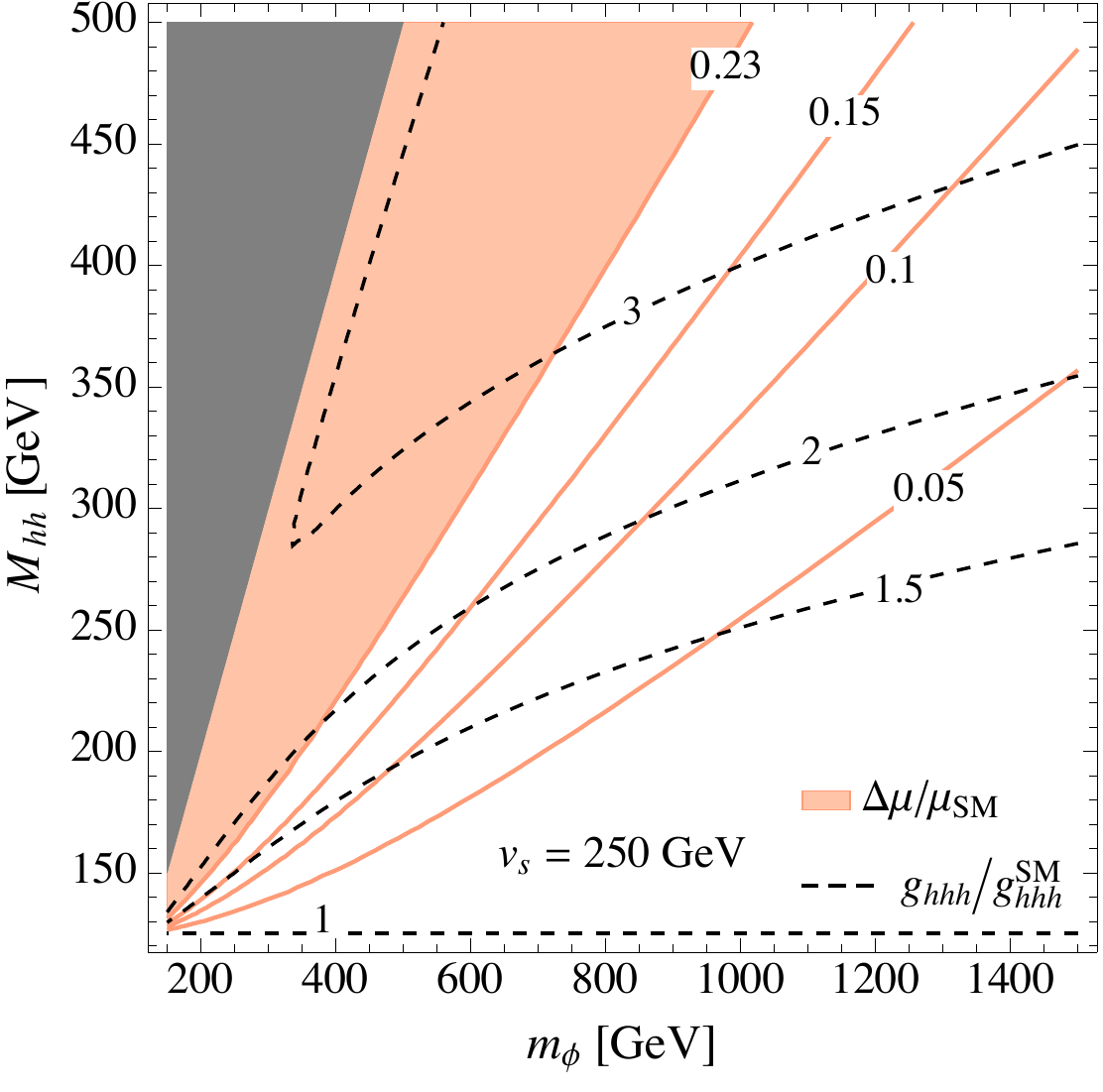}\hfill%
\includegraphics[width=.49\textwidth]{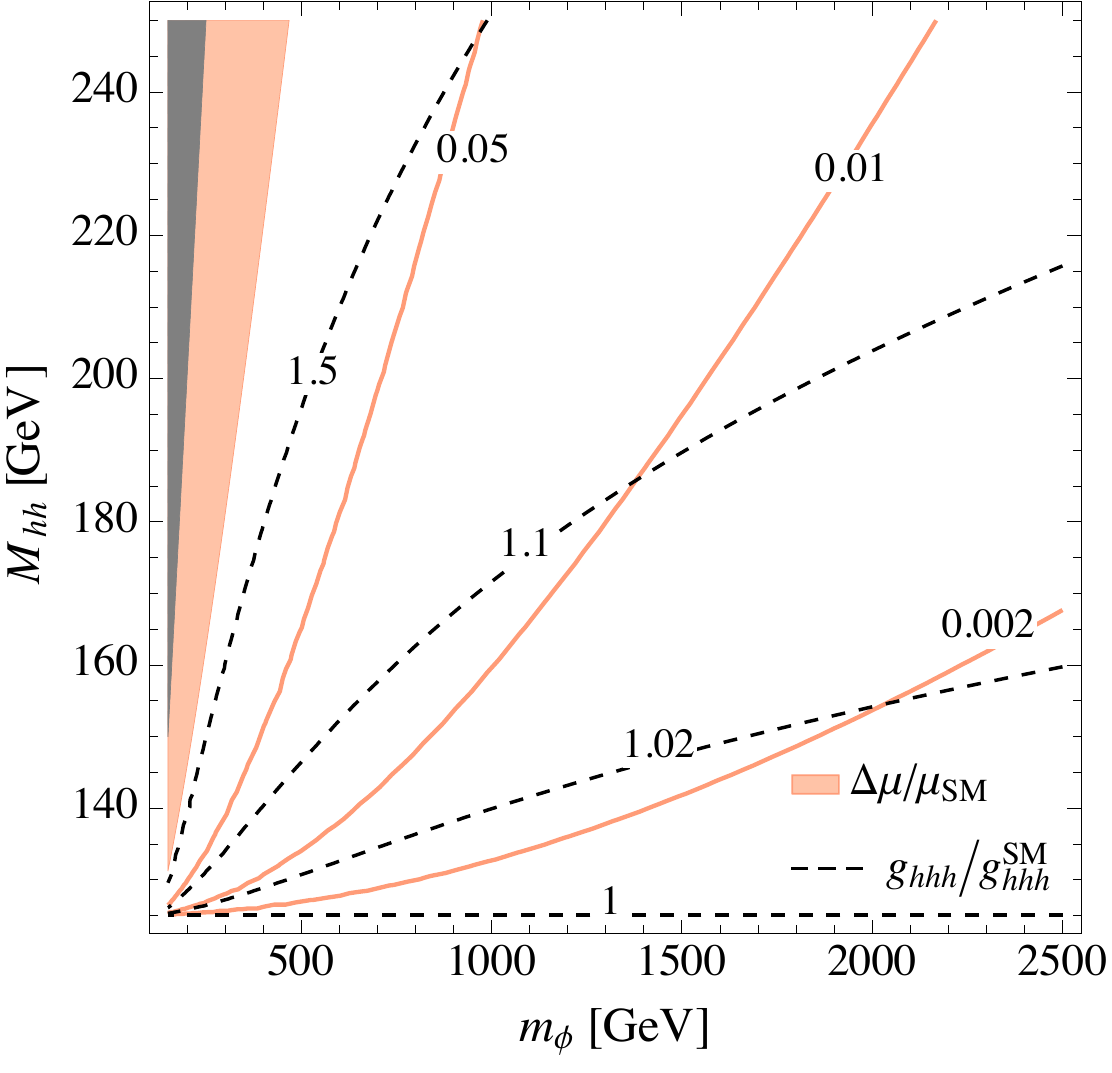}
\caption{\label{fig:triple1} Exclusion at 95\% C.L.\ from current Higgs couplings measurements at the LHC8 (coloured region), and deviation in the Higgs signal strengths (coloured lines). In both plots the ratio of the trilinear Higgs coupling to its SM value is drawn for $v_s = 250$ GeV (dashed black). Grey: unphysical parameters.}
\end{figure}

\begin{figure}[t]
\includegraphics[width=.48\textwidth]{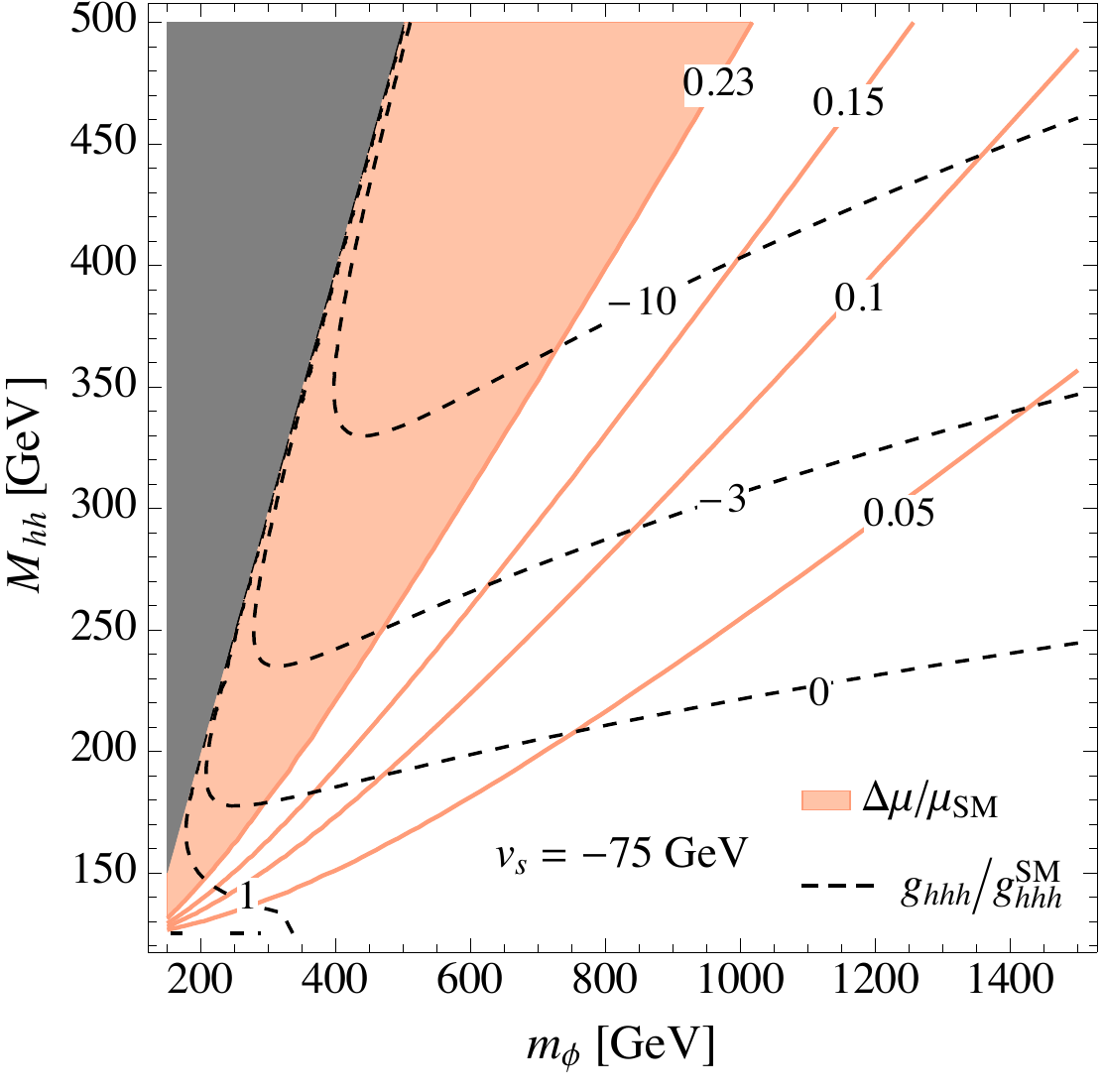}\hfill%
\includegraphics[width=.49\textwidth]{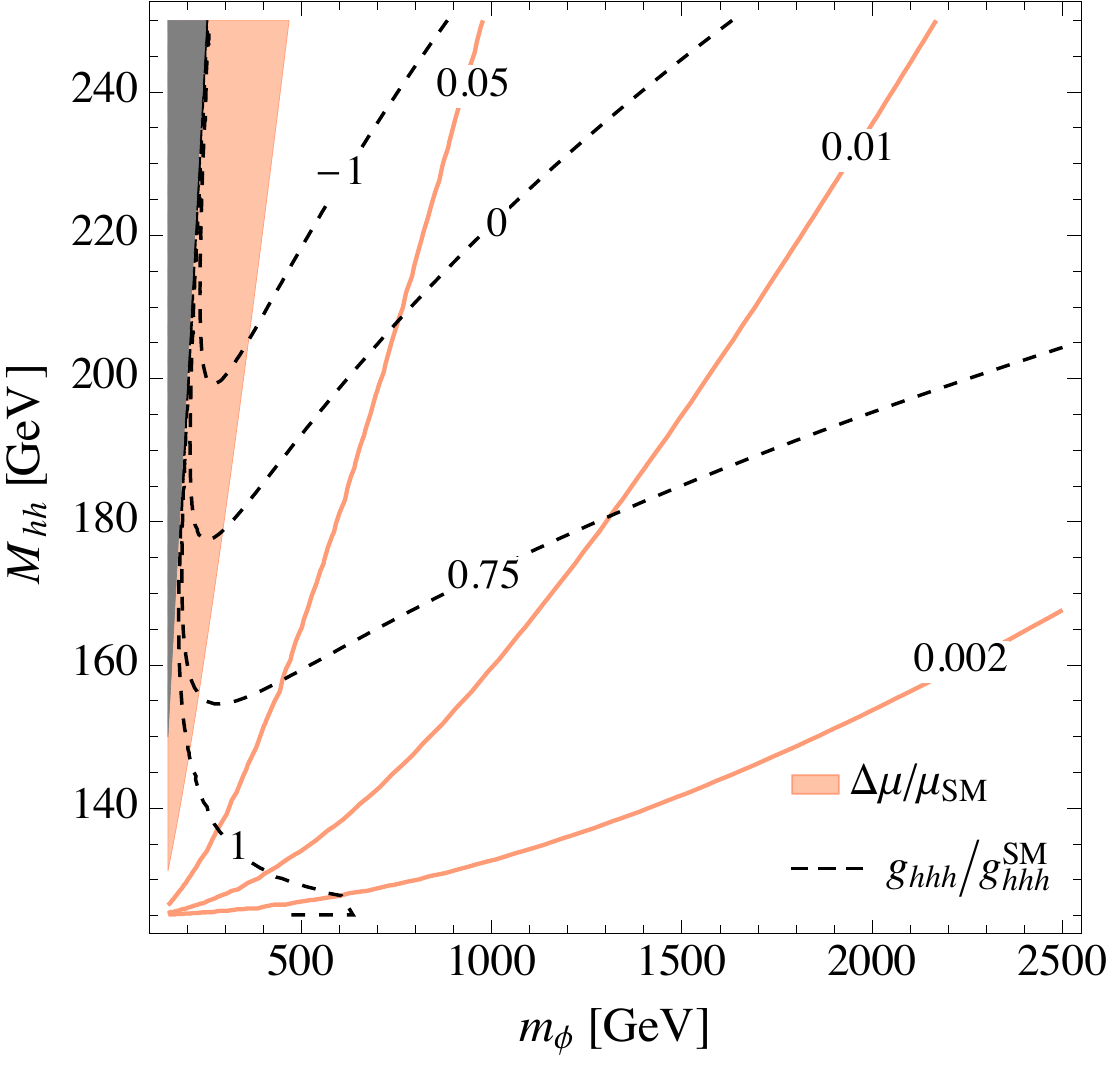}
\caption{\label{fig:triple2} Same as \figuresnames~\ref{fig:triple1}, but with $v_s = -75$ GeV.}
\end{figure}
\end{center}


\paragraph{Results and interplay of different measurements.}
The predictions for $s_\gamma^2$ and $g_{hhh}$ are shown in \figuresnames s~\ref{fig:triple1} and \ref{fig:triple2}. The shaded pink region is excluded by a combined fit to Higgs coupling measurements \cite{Giardino:2013bma,Falkowski:2013dza}, while continuous pink lines display some values of $s_\gamma^2$.
Given that the estimates for future sensitivities are only indicative, and that those for some machines (like FCC-hh) are not yet available, we prefer not to present lines corresponding to specific future colliders. This will make it easier to visualise the reach due both to new precision assessments, and to new data.
With this in mind and to avoid clutter, in the left-hand plots we focus on a region of parameters interesting for LHC and its upgrades, and in the right-hand ones we zoom into an area more interesting for future leptonic machines. One can see that reaching a per-mille precision on the Higgs signal strengths would constitute an impressive probe of this scenario. This will become even clearer when compared with direct searches in \sectionname~\ref{sec:comparison}.

We show the predictions for $g_{hhh}/g_{hhh}^{\rm SM}$ as dashed black lines. As discussed in the previous section, this quantity is not univocally fixed in terms of $m_\phi$ and $M_{hh}$, so we show it for the benchmark values $v_s = -75$ GeV and $v_s = 250$ GeV.
One can see that the allowed parameter space can accommodate very large deviations from one in $g_{hhh}/g_{hhh}^{\rm SM}$, with both signs and up to an order of magnitude. From \eqref{ghhh} one infers that the large size of such deviations is quite general, in other words it is not typical of the particular values chosen for $v_s$, at least until $v/v_s$ does not become too small. We explicitly verified this by exploring other values of $v_s$ not too far from $v$.
We further check numerically that variations of both $\lambda_{HS}$ and $\lambda_S$ have a minor impact in the size of $g_{hhh}/g_{hhh}^{\rm SM}$, confirming the insight from our analytical formula \eqref{ghhh} (the plots being for $\lambda_{HS} = \lambda_S = 1$).
Finally, another check we perform is that, in the whole interval of parameters that we consider, one never has $\Gamma_\phi > m_\phi$. This would be the case {\it e.g.}\ if we have chosen a value of $v_s$ closer to zero.

An important message of \figuresnames s~\ref{fig:triple1} and \ref{fig:triple2} and table~\ref{tab:Higgs_couplings} is that, at each stage of the future experimental program (except for FCC-ee and CEPC), the trilinear Higgs coupling and the Higgs signal strengths will be both crucially important as probes of the model.
Moreover, under some circumstances, $g_{hhh}$ could even be the only observable where deviations will show up.\footnote{This can be expected in a bottom-up perspective from the EFT analysis of \cite{Pomarol:2013zra}.}
Perhaps more important than that, in the currently allowed parameter space one could have deviations which are so large, to induce an increase of an order of magnitude (or more) of double Higgs production (see {\it e.g.}\ \cite{Baglio:2012np}), likely already in the reach of current machines.

The presence of so large deviations in the trilinear Higgs coupling is a rather peculiar property: the opposite is in fact true in many other models (see {\it e.g.}\ \cite{Gupta:2013zza}), for example when the Higgs mixes with another doublet. Our study shows an explicit and rather general model where such large deviations are actually typical. It then motivates a further effort in the community to measure the Higgs self-coupling, already with the available data, and to assess the relative sensitivity of future machines. 

\paragraph{Electroweak precision tests.}
In the model under consideration the only contribution to the electro-weak precision tests (EWPT) is due to the IR-logarithms in the $\hat{S}$ and $\hat{T}$ parameters, that can be estimated as \cite{Barbieri:2007bh}
\begin{equation}
\hat S =  \frac{\alpha}{48\,\pi\,s_w^2}\, s^2_\gamma \, \log \frac{m_\phi^2}{m_h^2} ,\qquad \hat T =  - \frac{3\alpha}{16\,\pi\,c_w^2}\, s^2_\gamma \, \log \frac{m_\phi^2}{m_h^2},
\label{EWPT}
\end{equation}
where $\alpha$ is the fine-structure constant and $s_w$ and $c_w$ are the sine and cosine of the Weinberg angle.
$\hat S$ and $\hat T$ are proportional to the reduction of the $h$ couplings to the weak bosons, {\it i.e.}\ to the mixing angle. They also vanish when $m_h=m_\phi$, being the mixing unphysical in this limit.
EWPT do not have a strong impact on the picture under consideration, unless $s_\gamma^2$ does not go to zero when $m_\phi$ is decoupled. For this to be possible, one needs $M_{hh}$ to be proportional to $m_\phi$ (see \eqref{singamma}): while this is not the case in Supersymmetry, Twin and Composite Higgs models realise such a behaviour. We will then return on the EWPT in \sectionname~\ref{sec:CHM_EWPT}.


\section{Direct searches}\label{sec:DirectSearches}

Searches for scalar resonances decaying into a pair of SM particles have been performed by the ATLAS and CMS experiments in a large variety of channels. Being based on the full dataset of the first LHC run, many of these searches are becoming sensitive to cross-sections that are significantly smaller than the ones of a standard Higgs boson. Therefore they already set relevant constraints for scalar singlets. With the increasing center-of-mass energy and luminosity, all these bounds will strengthen further. In this section we discuss the searches that are relevant for our purposes, and present extrapolations of their limits to future colliders.


\subsection{Extrapolation of the bounds}\label{sec:method}
In order to get a projection of the mass and cross-section reach of future collider experiments, we extrapolate the expected limits from the 8 TeV LHC relying on a parton luminosity rescaling, as introduced in \cite{SalamWeiler} and developed in \cite{Thamm:2015zwa}. This method, which we are going to recall here, is subject to a number of assumptions that are potentially not satisfied in a generic search. It nevertheless constitutes a quick and efficient tool to get a reasonable estimate of the masses and cross-sections that could be probed, avoiding to perform a full Monte-Carlo simulation of the experimental analysis.

The main underlying idea is that the number of signal events that can be excluded with a certain confidence level at a given collider, for a given mass of the resonance, depends exclusively on the corresponding number of background events $N_B$. In other words, starting from the LHC bounds, it is possible to obtain a projection for the exclusions at a different collider, as a function of the resonance mass, studying the scaling of the background with the collider energy and luminosity.


\paragraph{Extrapolation procedure.}
Concretely, one proceeds as follows \cite{Thamm:2015zwa}. Take a value of the mass, $m_0$, and a corresponding value of the cross-section times branching ratio, $[\sigma\times{\rm BR}]_0$, that can be excluded at 95\% C.L. at a collider with center-of-mass energy $\sqrt{s_0}$ and integrated luminosity $L_0$. It  is possible to find a new value of the mass, $m$, for which the same number of background events is produced at a different collider with energy $\sqrt{s}$ and luminosity $L$, by solving
\begin{equation}\label{bkg_equality}
N_B(m;s,L) = N_B(m_0;s_0,L_0).
\end{equation}
The number of signal events that can be excluded for the new mass $m$ with the new collider will then simply be equal to the one of the original search, $N_S(m;s,L) = N_S(m_0;s_0,L_0)$.

If the signal acceptances and efficiencies do not change significantly between the two experiments -- an assumption which we shall discuss further below -- one can finally get the new value of the excluded cross-section dividing the number of signal events by the luminosity $L$, and therefore, imposing \eqref{bkg_equality},
\begin{equation}\label{cross_section}
[\sigma\times {\rm BR}](m;s,L) = \frac{L_0}{L}[\sigma\times {\rm BR}]_0(m_0;s_0,L_0).
\end{equation}

The extrapolation of the backgrounds to higher energies and luminosities, which is needed to determine the mass $m$ from \eqref{bkg_equality}, is performed in the following way. The number of background events of a certain process can be written as
\begin{equation}\label{background}
N_B(m;s,L) \propto L\cdot \sum_{\{i,j\}}\int \d\hat s \frac{\d\L_{ij}(s,\hat s)}{\d\hat s}\hat\sigma_{ij}(\hat s),
\end{equation}
where $\d\L_{ij}(s,\hat s)/\d\hat s$ is the parton luminosity for a channel involving two initial partons $i, j$ with a center-of-mass energy $\sqrt{\hat s}$, and $\hat\sigma_{ij}(\hat s)$ is the partonic cross-section for that process. The sum is performed over all the partonic reactions that contribute to the background. An explicit expression for the parton luminosities in terms of the parton distribution functions is given in \eqref{PDF} in the appendix.

Two simplifications can now be made. We assume the resonance to be narrow, hence the integral in \eqref{background} is performed over a small energy range $\Delta\hat s$ close to the resonance peak, with $\Delta\hat s \ll m^2$. The integrand can then be taken approximately constant over the integration range. If furthermore one performs the searches in a window of constant relative width $\Delta\hat s/m^2$, the equality \eqref{bkg_equality} simplifies to
\begin{equation}
m^2 \sum_{\{i,j\}}\frac{\d\L_{ij}}{\d \hat s}(s,m^2)\hat\sigma_{ij}(m^2) = m_0^2 \frac{L_0}{L} \sum_{\{i,j\}}\frac{\d\L_{ij}}{\d \hat s}(s_0,m_0^2)\hat\sigma_{ij}(m_0^2).
\end{equation}

At energies much higher than the SM thresholds, the cross-sections $\hat \sigma_{ij}$ decrease as $\hat s^{-1}$, and one can finally write
\begin{equation}\label{final_bkg}
\sum_{\{i,j\}}c_{ij} \frac{\d\L_{ij}}{\d \hat s}(s,m^2) = \frac{L_0}{L}\sum_{\{i,j\}}c_{ij} \frac{\d\L_{ij}}{\d \hat s}(s_0,m_0^2),
\end{equation}
where the coefficients
$c_{ij} \equiv \lim_{\hat s\to\infty}\hat s\hat \sigma_{ij}(\hat s)$ determine the composition of the background. While the $\hat s^{-1}$ behaviour of the cross-sections is clearly a good approximation for high-mass resonances, as was the case in \cite{Thamm:2015zwa}, one may wonder whether it can be used also in our context, where one is interested in scalar particles that are not too much heavier than the Higgs boson. We verified that \eqref{final_bkg} is actually valid down to a few hundreds of GeV, depending on the process. In any case, we do not use the experimental constraints below the region of validity for our extrapolations, and we do not provide extrapolated exclusions for masses below that limit. More details on this issue are given in appendix~\ref{sec:appendix}.

Since the extrapolated values of $m$ obtained from \eqref{final_bkg} are higher than the original $m_0$ for a collider with increased energy or luminosity, the range of masses where one gets a projection of the bounds will be shifted to higher values. Often, however, the physically more interesting region -- at or below the TeV scale-- lies outside that range, especially in the case of the colliders with highest luminosity or energy.
In order to get an extrapolated exclusion in the full region of interest, in \cite{Thamm:2015zwa} the following method was used. One notices that, for a given energy of the new collider, the values of the masses where an exclusion is provided grows with the luminosity $L$. This follows immediately from \eqref{final_bkg} and from the fact that the parton luminosities are a decreasing function of $\hat s$. Then, one can easily extend the extrapolation of the bounds to lower masses, decreasing the luminosity of the collider from $L$ to $L'<L$; lowering the luminosity enough, one will eventually be able to get an exclusion for arbitrary masses down to the initial values. However, since the excluded cross-section increases with the inverse of $L'$, the bounds for low masses will be much less powerful, and can be interpreted only as upper bounds on the actual reach of a collider. This is what has been done in \cite{Thamm:2015zwa}.

In order to get more realistic projections for low masses, in our analysis we rescale the cross-section limits, for each value of the mass, by the square root of the ratio between the nominal $L$ of the new collider and the luminosity that was actually used for the extrapolation, $L'$. This is a quite accurate approximation of the scaling of the excluded cross-section with the luminosity when the systematic errors are small (see \eqref{significance}).

To summarise, starting from an LHC exclusion $[\sigma\times{\rm BR}]_0$, the corresponding projection for a different collider, for a given mass $m$, reads
\begin{equation}
\lbrack\sigma\times{\rm BR}\rbrack(m;s,L) = \min_{L' \leq L}\left[\frac{L_0}{\sqrt{L L'}}\lbrack\sigma\times{\rm BR}\rbrack_0(m_0;s_0,L_0)\Big|_{m_0(L')}\right],
\end{equation}
where $m_0(L')$ is determined from \eqref{final_bkg}, substituting $L$ with $L'$.


\paragraph{Discussion of the main assumptions.}
Let us finally make a few comments on our prime assumption that the exclusion is driven by the background alone, which is not necessarily very accurate. First of all, the composition of the background must not change dramatically between the two experiments: the background that is dominant at the LHC must still be the leading one at higher energies. Next, the assumption is justified only if the analysis is done as a counting experiment of signal vs.\ background events, assuming the resonance to be narrow. If effects due to the shape of the resonance peak are taken into account, the results depend on signal and background kinematical distributions in a non-trivial way.\footnote{This is however what is done by the experimental collaborations at the LHC; we assume that the results do not deviate much from those obtained with a cut-and-count analysis.} We always work in the narrow width approximation.
The significance of a signal over a given background then is
\begin{equation}\label{significance}
\mathrm{significance} = \frac{N_S}{\sqrt{N_B + \alpha^2 N_B^2 + \beta^2 N_S^2}},
\end{equation}
where $\alpha$ and $\beta$ are the systematical errors on the background and signal, respectively. In the hypothesis that the background dominates over the signal, which is always verified in our case, this last error can safely be neglected. However, in order to have the same significance at different collider experiments when $N_B$ is fixed, one must assume that the systematic error $\alpha$, when it is not negligible, remains approximately constant.

A further issue is represented by the fact that the values of $N_S$ and $N_B$ in \eqref{significance} in general depend on the acceptances and efficiencies, due to the experimental analysis and detector effects. They could vary for different experiments, especially when extrapolating over a wide range of energies -- {\it e.g.}\ from 8 TeV to 100 TeV. We will nevertheless assume, following \cite{Thamm:2015zwa}, all the acceptances and efficiencies to be independent of the collider energy and the resonance mass.
This assumption is another possible limitation of the method, but it is quite reasonable for the second stage of the LHC. For higher energy colliders, our results should be viewed as an estimate of what limits could be achieved assuming an experimental setup similar to the present one, and at the same time as a goal for the future detectors.

\begin{table}[t]
\renewcommand{\arraystretch}{1.2}
\begin{center}
\small
\begin{tabular}{r||cc|cc}
& \multicolumn{2}{c}{ATLAS} & \multicolumn{2}{c}{CMS} \\ \hline
$ZZ(4\ell)$ & $\sqrt{s}=7-8~\mathrm{TeV}$, $L=4.6+20.7/\mathrm{fb}$ & \cite{ATLAS:2013nma} & $\sqrt{s}=7-8~\mathrm{TeV}$, $L=5.1+19.7/\mathrm{fb}$& \cite{Chatrchyan:2013mxa}\\
$ZZ(2\ell 2\nu)$ & $\sqrt{s}=7~\mathrm{TeV}$, $L=4.7/\mathrm{fb}$  & \cite{Aad:2012ora} & '' & \cite{Khachatryan:2015cwa}\\
$ZZ(2\ell 2j)$ & ''  & \cite{Aad:2012oxa} & '' & \cite{Khachatryan:2015cwa}\\
$ZZ(2\ell 2\tau)$ &---  &  & '' & \cite{Khachatryan:2015cwa}\\
$WW(2\ell2\nu)$ & $\sqrt{s}=8~\mathrm{TeV}$, $L=21/\mathrm{fb}$ & \cite{ATLAS-CONF-2013-067} & '' & \cite{Khachatryan:2015cwa}\\
$WW(\ell\nu jj)$ & $\sqrt{s}=7~\mathrm{TeV}$, $L=4.7/\mathrm{fb}$ & \cite{Aad:2012me} & $\sqrt{s}=8~\mathrm{TeV}$, $L=19.3/\mathrm{fb}$& \cite{CMS-PAS-HIG-13-027}\\[4pt]

$VV ({\rm all})$ & --- & &$\sqrt{s}=7-8~\mathrm{TeV}$, $L=5.1+19.7/\mathrm{fb}$ & \cite{Khachatryan:2015cwa}\\
\hline
$hh(4b)$ & $\sqrt{s}=8~\mathrm{TeV}$, $L=19.5/\mathrm{fb}$&\cite{ATLAS-CONF-2014-005} & $\sqrt{s}=8~\mathrm{TeV}$, $L=17.9/\mathrm{fb}$ & \cite{Khachatryan:2015yea}\\
$hh(bb\gamma\gamma)$ & $\sqrt{s}=8~\mathrm{TeV}$, $L=20/\mathrm{fb}$&\cite{Aad:2014yja} &$\sqrt{s}=8~\mathrm{TeV}$, $L=19.7/\mathrm{fb}$&\cite{CMS-PAS-HIG-13-032}
\end{tabular}
\caption{\label{tab:direct-searches}\small Direct searches relevant for our analysis performed at the LHC experiments.}
\end{center}
\end{table}


\subsection{Current and future constraints}
Given that we are interested in models with a scalar singlet, the direct searches relevant to us are those for a heavy scalar decaying into $W^+W^-$, $ZZ$ and $hh$ final states. The summary of all the searches that we have considered is provided in table~\ref{tab:direct-searches}. These are only a partial list of all the searches for extra scalars carried out by LHC experiments, but they are the most sensitive ones for scenarios where the extra scalar is a singlet.

Please note that all the extrapolations that we present here are valid for a generic spin-0 resonance decaying into $ZZ$, $WW$ and $hh$, and have therefore a general validity beyond the singlet scenario that we consider in this paper.


\paragraph{The {\em ZZ} and {\em WW} channels.}
\begin{figure}[t]
\centering%
\includegraphics[width=0.49\textwidth]{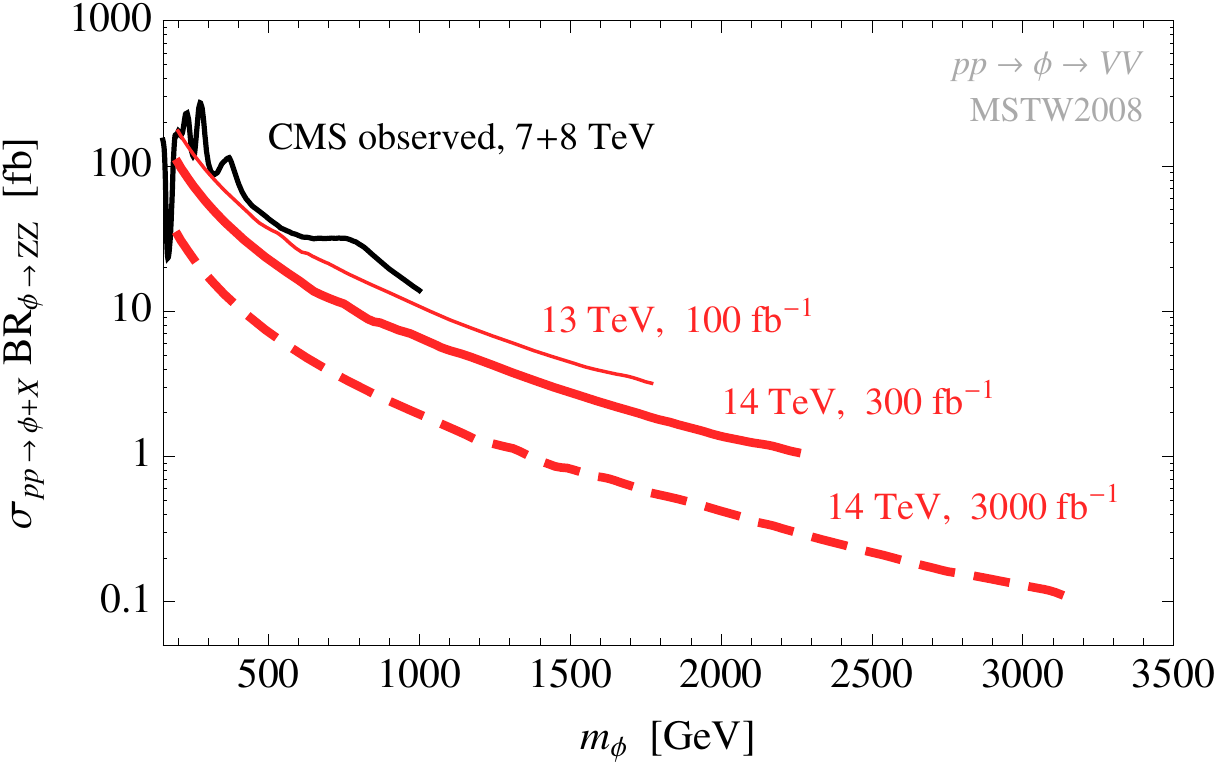}\hfill%
\includegraphics[width=0.49\textwidth]{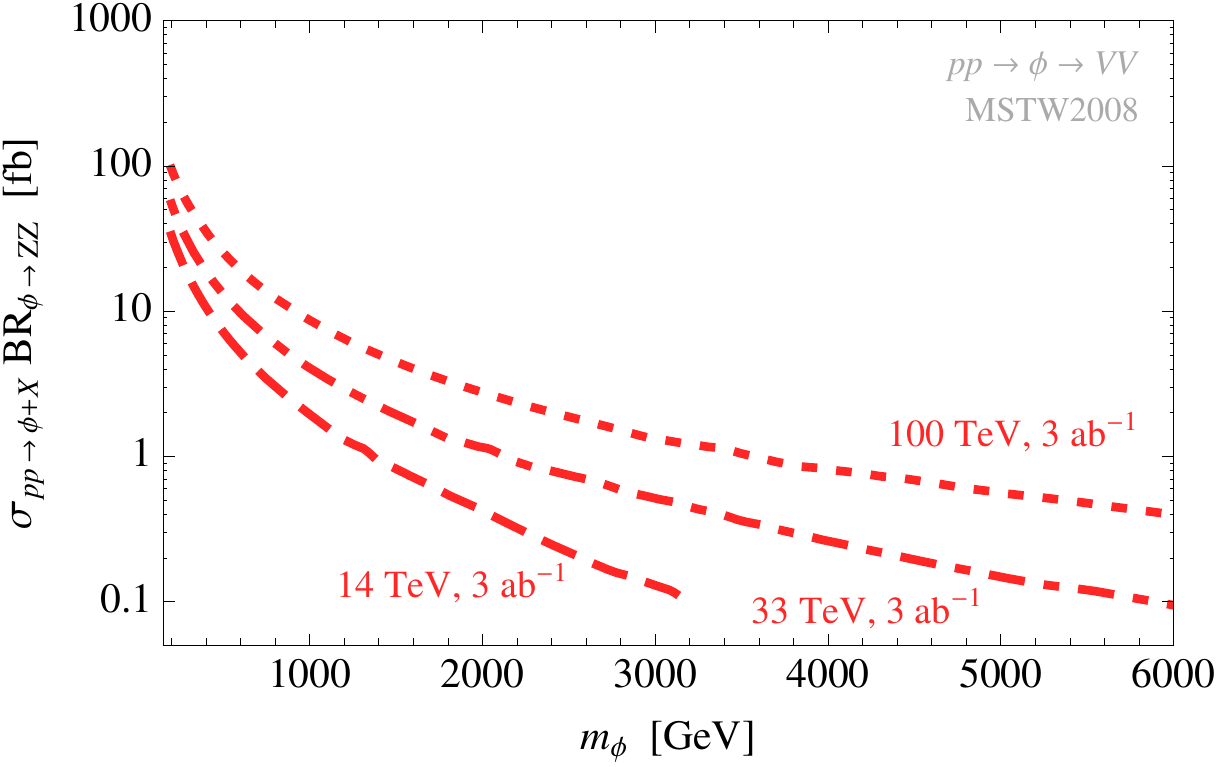}
\caption{Left: CMS 95\% C.L. exclusion (black) on $\sigma_{pp\to\phi}\times{\rm BR}_{\phi\to ZZ}$ from a combination of the $ZZ$ and $WW$ channels, together with the projections for the LHC at 13 TeV with 100 fb$^{-1}$ (thin red), and at 14 TeV with 300 fb$^{-1}$ (solid red) and 3000 fb$^{-1}$ (red dashed). Right: reach in the $hh\to VV$ channel of the HL-LHC (dashed), HE-LHC (dot-dashed) and FCC-hh (dotted) with 3 ab$^{-1}$.
\label{fig:VVextrapolations}}
\end{figure}
Many different searches for scalar resonances decaying into $WW$ and $ZZ$ in fully leptonic and semileptonic final states have been presented both by ATLAS and CMS and are summarised in table~\ref{tab:direct-searches}. The strongest individual constraint comes always from the $ZZ\to 4\ell$ and $ZZ\to 2\ell 2\nu$ channels, which dominate respectively below and above about 500 GeV. The $WW$ and the other $ZZ$ channels give a bound on the cross-section which is always weaker by a factor of a few. However, a recent analysis by CMS \cite{Khachatryan:2015cwa} shows that the inclusion of all these final states is able to significantly improve the limit over all the mass range.\footnote{In the same paper, the results are explicitly interpreted in the case of a scalar singlet extension of the SM.} We will use only this global combination for our analysis in the di-vector channel.

The background in the $ZZ\to 4\ell$ channel, which dominates the exclusion in the region of interest for our extrapolation, comes to a large extent from non-resonant $ZZ$ production, and is mainly produced through the $q\bar q\to ZZ$ partonic reaction \cite{Chatrchyan:2013mxa}.
We therefore extrapolate this background using the $\L_{q\bar q}$ parton luminosities, using everywhere the MSTW2008 parton distribution functions \cite{Martin:2009iq}.

The experimental exclusion is provided in the mass range from 145 GeV up to 1 TeV, but we consider only masses above 200 GeV because below that value the SM thresholds influence the shape of the background and the extrapolation procedure breaks down. We explicitly checked that above 200 GeV the excluded cross-section is driven by the $\L_{d\bar d}$ parton luminosity to a very good approximation (see appendix~\ref{sec:appendix}).

In \figuresnames~\ref{fig:VVextrapolations} we show our results for the projected limits on $\sigma_{pp\to\phi}\times{\rm BR}_{\phi \to ZZ}$. The left-hand panel shows the projections for the LHC at 13 TeV with an integrated luminosity of 100 fb$^{-1}$, and at 14 TeV with a luminosity of 300 fb$^{-1}$ and 3000 fb$^{-1}$, together with the current constraint from CMS. The right-hand plot shows the extrapolations for the luminosity and energy upgrades of the LHC, respectively with a center-of-mass energy of 14 TeV and 33 TeV and both with a luminosity of 3 ab$^{-1}$,
and for the 100 TeV FCC-hh collider, also considering a luminosity of 3 ab$^{-1}$. The reach of FCC-hh extends up to masses of about 11 TeV.

We have checked that our limits are in very reasonable agreement with those obtained in the Snowmass study \cite{Brownson:2013lka}, where a full collider simulation of the signal and the background has been carried out, and with the ones for the HL-LHC presented by CMS in \cite{CMS-PAS-FTR-13-024}.

One sees that the next phase of the LHC will improve the present bounds by more than a factor of 2 -- with some improvement to be expected already in the first 13 TeV run -- while another factor of 3 will be achieved by the high luminosity phase. In terms of the absolute limit on the cross-section, the high-energy colliders HE-LHC and FCC-hh will give weaker constraints, but one has to keep in mind that the cross-sections for a given mass $m_\phi$ will be much larger at such colliders.


\paragraph{The {\em hh} channels.}
Searches for resonant $hh$ production have been performed by both the ATLAS and CMS collaborations. The two most sensitive channels at present are $hh\to 4b$ \cite{ATLAS-CONF-2014-005,Khachatryan:2015yea}, which gives the strongest exclusions for masses $m_\phi > 400$ GeV, and $hh\to 2b\,2\gamma$ \cite{Aad:2014yja,CMS-PAS-HIG-13-032}, which is more constraining at low masses. Other decay channels, {\it e.g.}\ involving $\tau$ leptons, could also become relevant for the next LHC run and future experiments \cite{No:2013wsa,Martin-Lozano:2015dja,Kotwal:2015rba}.

The background in the $4b$ channel is dominated by $t\bar t$ and multijet events \cite{Khachatryan:2015yea}. In terms of partonic processes, both $gg$ and $q\bar q$ contribute, but the gluon-fusion background is expected to dominate at high energies, hence we will use the $\L_{gg}$ parton luminosities for the extrapolation. As in the previous case, we verified that the inclusion of the other parton luminosities would change the result only marginally. We have checked explicitly that the extrapolation procedure is robust above about $500$ GeV, where the $t\bar t$ threshold starts to influence the background (see appendix~\ref{sec:appendix}), and therefore we restrict the exclusion to masses above this value.

\begin{figure}[t]
\centering%
\includegraphics[width=0.49\textwidth]{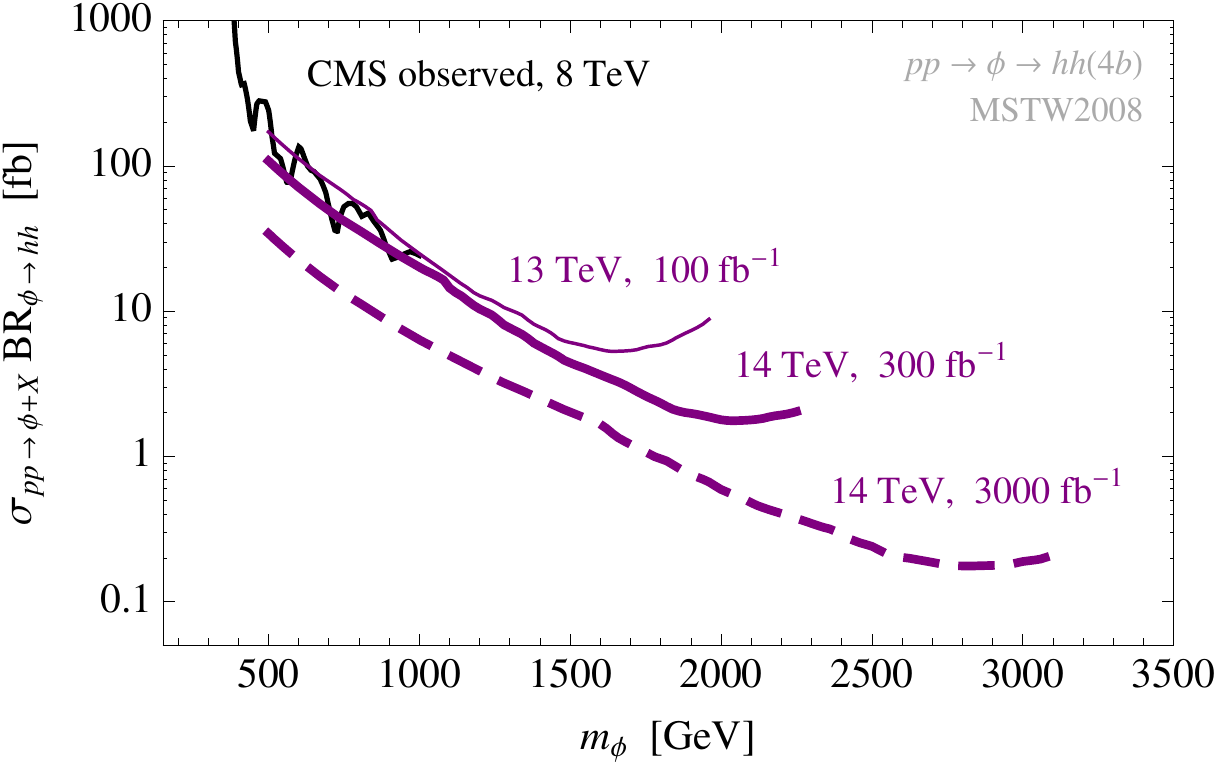}\hfill%
\includegraphics[width=0.49\textwidth]{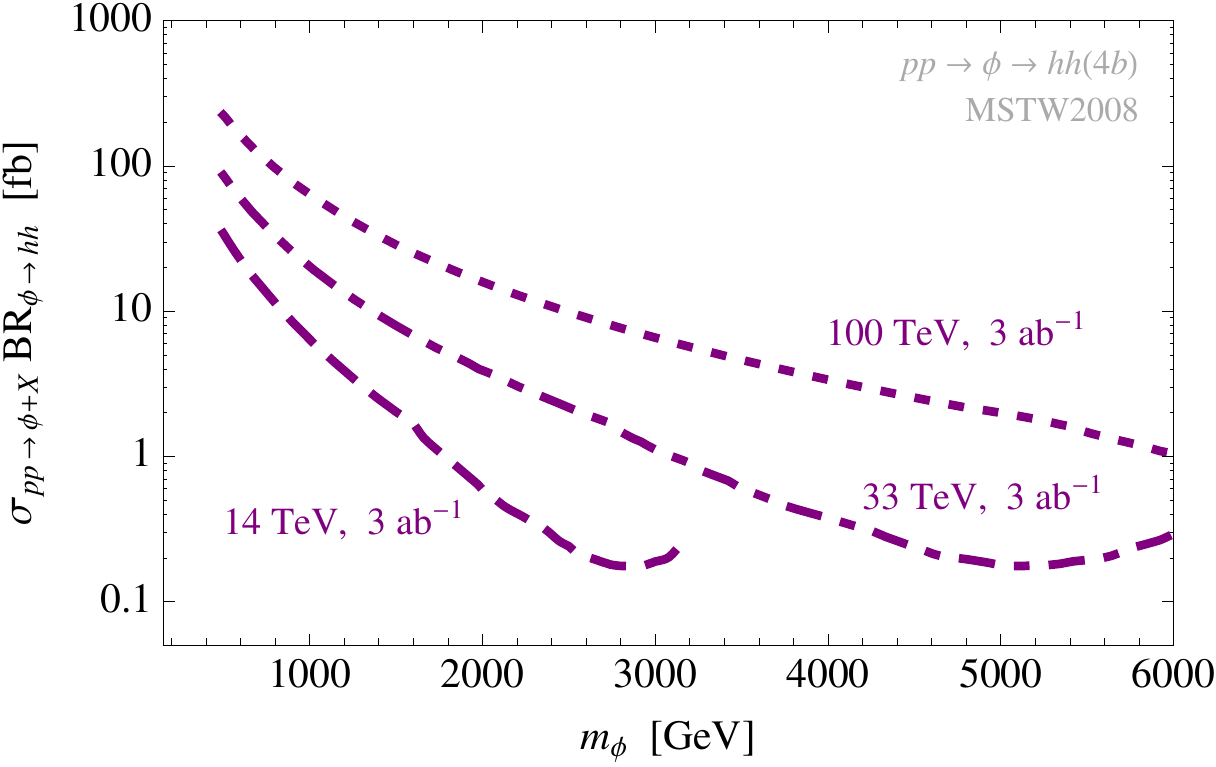}
\caption{Left: CMS 95\% C.L. exclusion (black) on $\sigma_{pp\to\phi}\times{\rm BR}_{\phi\to hh}$ from the $4b$ channel, together with the projections for the LHC at 13 TeV with 100 fb$^{-1}$ (thin purple), and at 14 TeV with 300 fb$^{-1}$ (solid purple) and 3000 fb$^{-1}$ (purple dashed). Right: reach in the $hh\to 4b$ channel of the HL-LHC (dashed), HE-LHC (dot-dashed) and FCC-hh (dotted) with 3 ab$^{-1}$.
\label{fig:4bextrapolations}}
\end{figure}

In \figuresnames~\ref{fig:4bextrapolations} we show our results for the projected limits on $\sigma_{pp\to\phi}\times{\rm BR}_{\phi \to hh}$ from the $4b$ channel, extrapolating the CMS exclusions \cite{Khachatryan:2015yea}. The energy and luminosity benchmarks and the notation are the same as in \figuresnames~\ref{fig:VVextrapolations}. The improvements from LHC8 to the future machines is roughly the same as for the $VV$ channel.

Concerning the $2b\, 2\gamma$ channel, its extrapolation is trustworthy only above 400 GeV, and in this region the $4b$ exclusion is always more powerful. For this reason we do not consider this channel for our future projections. We however include in our analysis the present exclusion which is relevant for low masses. We use also here the CMS data \cite{CMS-PAS-HIG-13-032}, since ATLAS sees a slight excess in this channel \cite{Aad:2014yja}.

\subsection{Results and comparison of bounds}\label{sec:comparison}

\begin{figure}[t]
\centering%
\includegraphics[width=.48\textwidth]{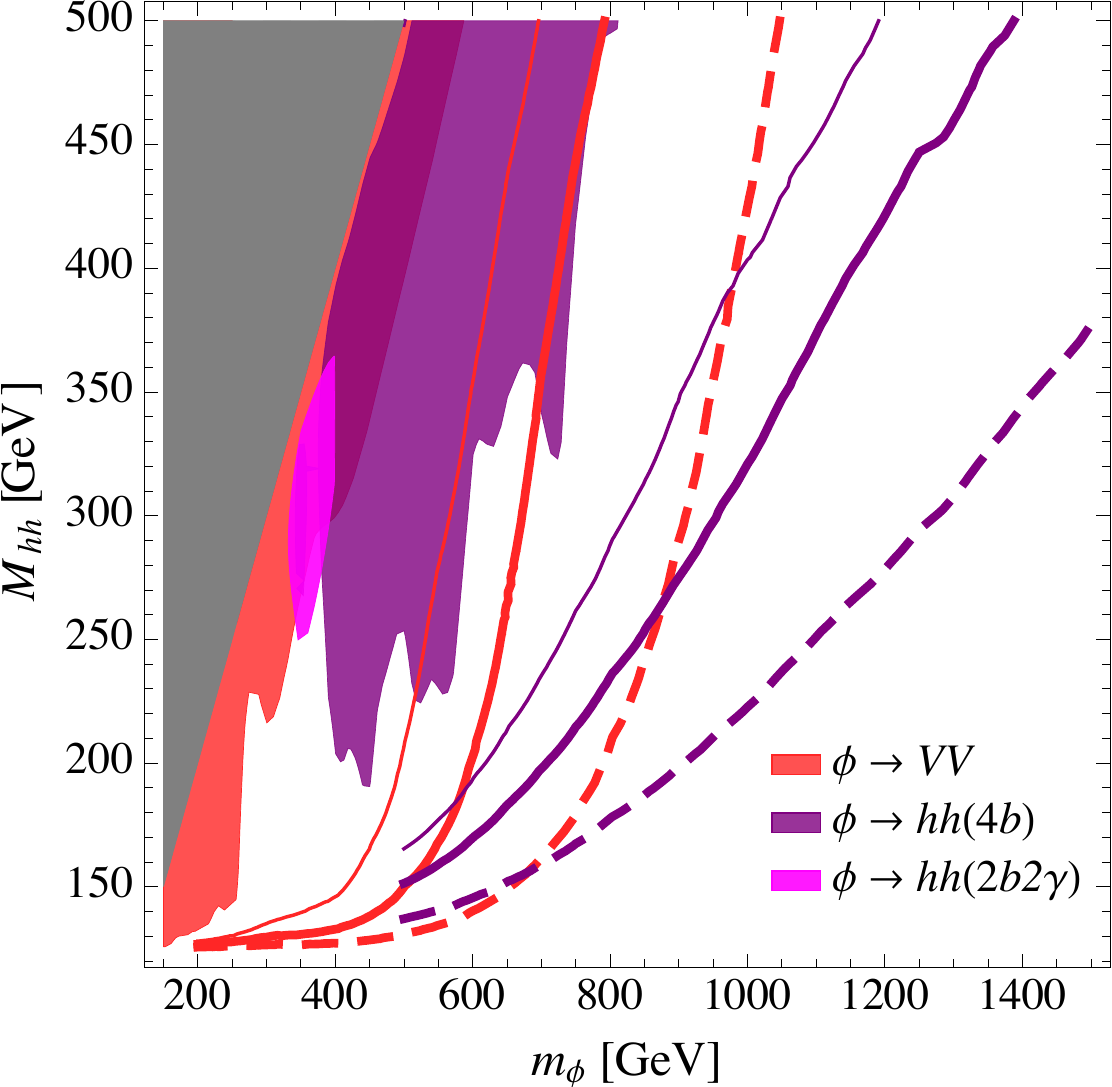}\hfill%
\includegraphics[width=.49\textwidth]{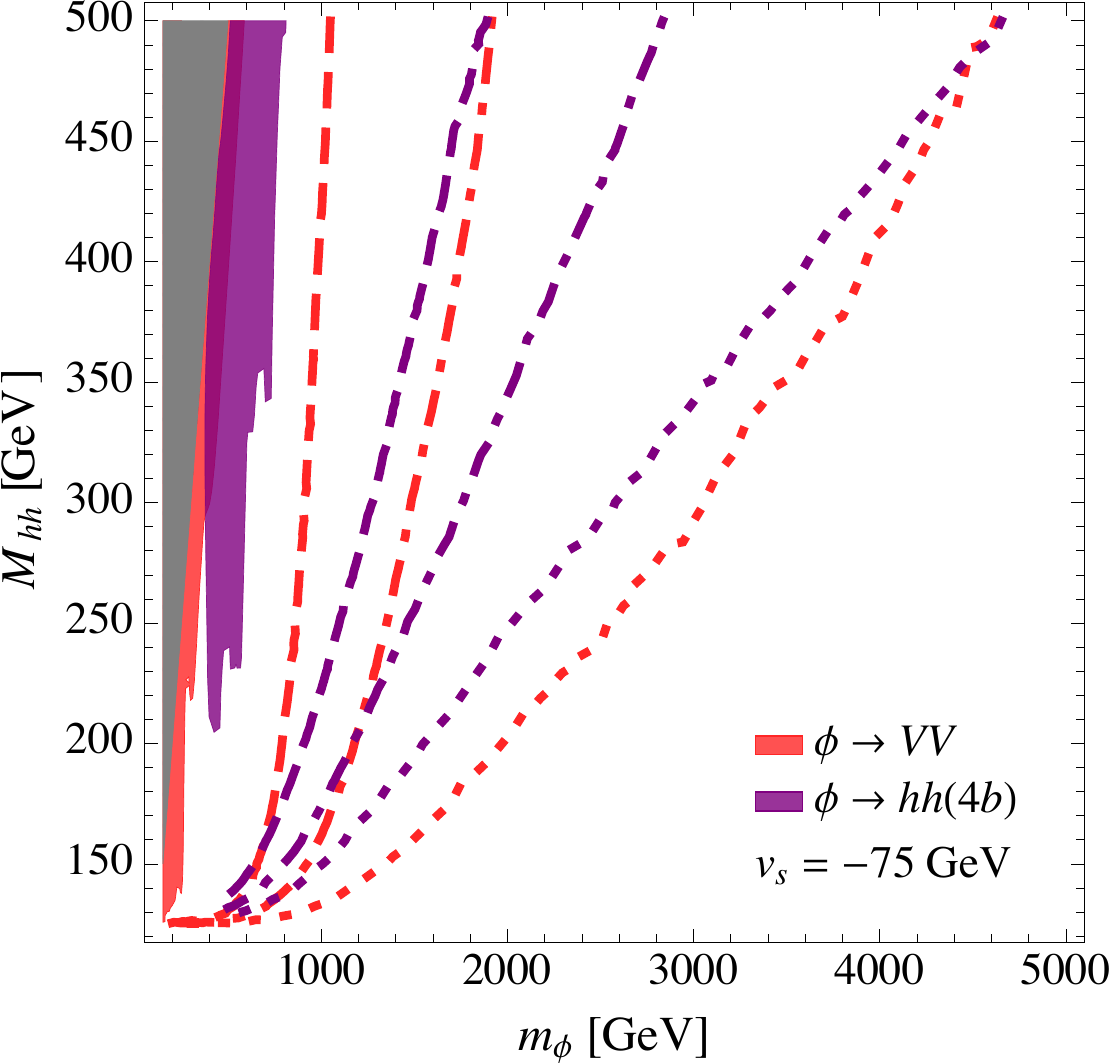}
\caption{Reach of the direct searches for $\phi\to VV$ (red), $\phi\to hh(4b)$ (purple), and $\phi\to hh(2b2\gamma)$ (magenta), for $v_s = -75$ GeV, and $\lambda_{HS} = \lambda_S = 1$. Grey: unphysical region. Left: current LHC exclusion at 95\% C.L.\ (coloured regions), and projections for LHC13 (thin lines), LHC14 (solid lines) and HL-LHC (dashed lines). Right: comparison with the projections for HE-LHC (dot-dashed lines) and FCC-hh (dotted lines).  \label{fig:direct75}}
\end{figure}

We now interpret the bounds presented in the previous section in the context of a generic scalar singlet extension of the SM.

The total production cross-section times branching ratio for $\phi$ is given by \eqref{eq:signal_strength} and \eqref{eq:signal_strength_hh}. When dealing with current bounds we always use the SM Higgs total cross-section and branching ratios given by the Higgs Cross Section Working Group \cite{Heinemeyer:2013tqa}, which are provided for masses up to 1 TeV. On the other hand, since the extrapolated limits extend up to much higher energies, where we cannot rely on the results of \cite{Heinemeyer:2013tqa}, we use the LO quantities in our projections. In particular, we compute the gluon-fusion cross-section with the program {\sc higlu} \cite{Spira:1995rr,Spira:1995mt}, and we use {\sc madgraph} \cite{Alwall:2011uj} to calculate the vector boson fusion and VH associated production cross-sections. The sum of these three components constitutes a good approximation of the total Higgs production cross-section. Concerning the branching ratios, we use the results of \cite{Heinemeyer:2013tqa} for masses below 250 GeV, and we match them to the LO (plus NLO QCD) results above that scale, where the two quantities are practically equal.

\begin{figure}[t]
\centering%
\includegraphics[width=.48\textwidth]{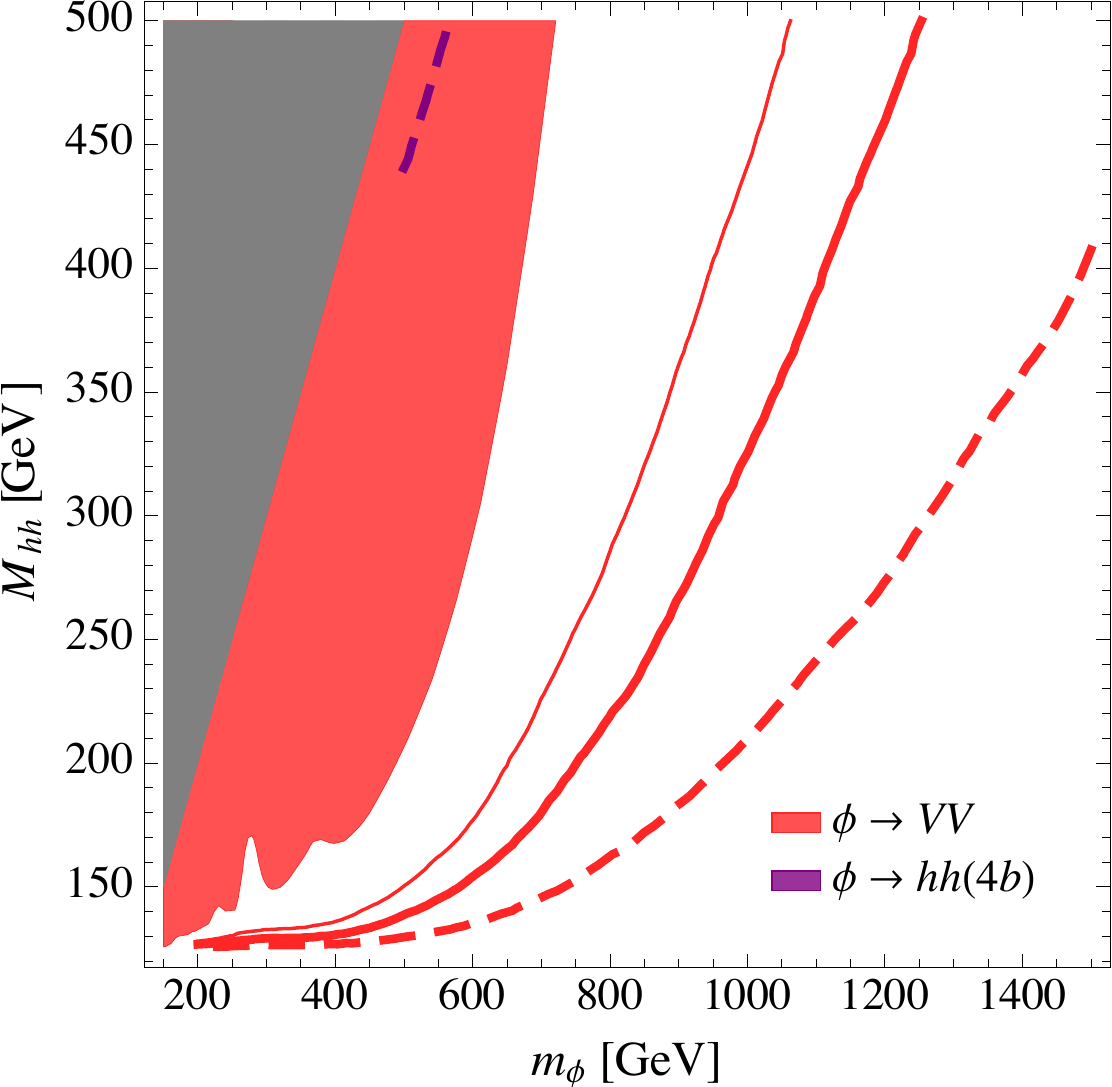}\hfill%
\includegraphics[width=.49\textwidth]{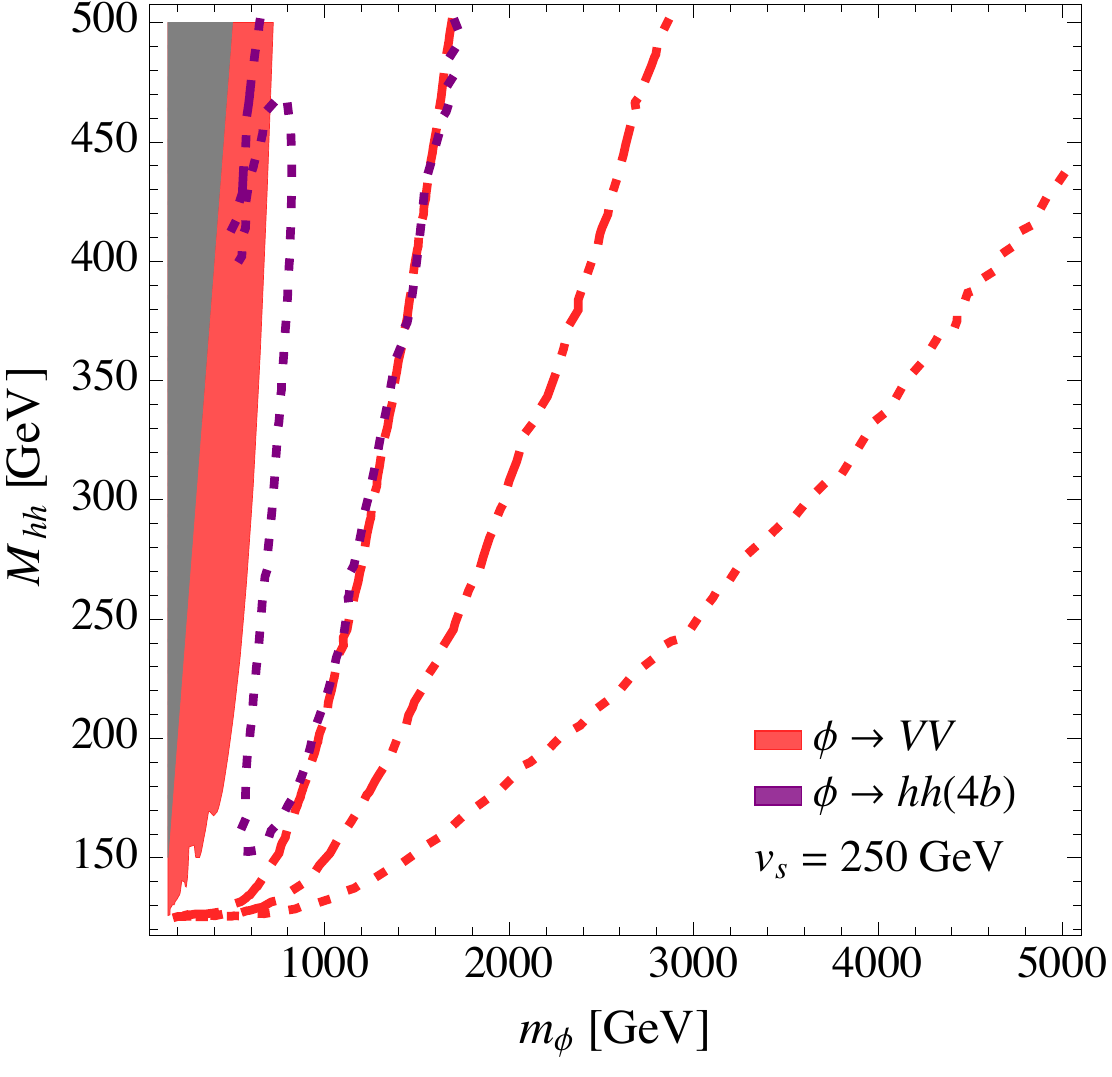}
\caption{Reach of the direct searches in the $VV$ (red) and $hh(4b)$ (purple) channels, for $v_s = 250$ GeV, and $\lambda_{HS} = \lambda_S = 1$. The notation is the same as in \figuresnames~\ref{fig:direct75}.
\label{fig:direct250}}
\end{figure}

In \figuresnames~\ref{fig:direct75} and \figuresnames~\ref{fig:direct250} we show the present and future exclusions in the same $m_\phi-M_{hh}$ plane of \figuresnames~\ref{fig:triple1}, for two different values of the singlet vev $v_s$. We fixed $\lambda_{HS} = \lambda_S =1$, however, as discussed above, the impact of their precise value on the results is very weak. The left-handed panels of both figures show the projections for the present and future stages of the LHC, up to the luminosity upgrade, while in the right-handed panels we show the range relevant for future high-energy colliders -- HE-LHC and the 100 TeV FCC-hh. The constraints from the $VV$ channel are drawn in red, and the ones from the two $hh(4b)$ and $hh(2b2\gamma)$ channels in purple and magenta, respectively. As discussed above, we do not show future projections for the $2b2\gamma$ channel, since they are never relevant in the region where we perform the extrapolations.

One sees immediately that, depending on the value of $v_s$, the relative importance of the $VV$ and $hh$ searches can be very different. In particular, we show the cases of $v_s = -75$ GeV and $v_s = 250$ GeV as benchmarks of a large enhancement and suppression of ${\rm BR}_{\phi\to hh}$, respectively.
In the first case the searches in the $hh$ channel dominate both at the 8 TeV and 14 TeV LHC, but for the larger masses that can be reached at future colliders their reach becomes comparable with the one from the $VV$ channel. In the second case of \figuresnames~\ref{fig:direct250}, on the other hand, ${\rm BR}_{\phi\to hh}$ is so strongly suppressed that searches in $hh(4b)$ will barely be sensitive to this scenario even at FCC-hh, while $VV$ always gives a much stronger exclusion.

The message that emerges from these figures is therefore pretty clear: at a high-energy collider, be it HE-LHC or FCC-hh, the $VV$ searches will always put the strongest bound on the production of a singlet-like scalar. On the contrary, at the various phases of the LHC the $hh\to 4b$ process can be the most powerful probe, given some particular choice of the parameters.
This is easily understood in terms of the fact that the enhancement (or suppression) of the $hh$ branching ratio is particularly effective at low masses, where the deviations from the limit \eqref{eq:BRasymptotic} can be larger (see \eqref{BRh2hh}), but in the region of parameter space relevant for future hadron colliders (high masses and small mixing angle) it is not possible to modify ${\rm BR}_{\phi\to hh}$ more than a certain amount.
Despite this strong dependence on the choice of parameters of the bounds in the $VV$ and $hh$ channels, the combined reach of the two exclusions is very similar in all cases: a suppression in ${\rm BR}_{\phi\to hh}$ will enhance the branching ratio into vectors, and vice versa, and the constraints on the cross-section in the two channels are of comparable order of magnitude.

We further point out that the enhancement of ${\rm BR}_{\phi\to hh}$ is correlated with the modification of the triple Higgs coupling, as one can see from \eqref{BRh2hh} and \eqref{ghhh}, and comparing \figuresnames~\ref{fig:triple2} with \figuresnames~\ref{fig:direct75}. This implies in particular that, if a resonance in the $hh$ channel will be discovered in the future, in the singlet scenario under consideration one expects to see an enhancement also in non-resonant double Higgs production.

\begin{figure}[t]
\centering%
\includegraphics[width=.48\textwidth]{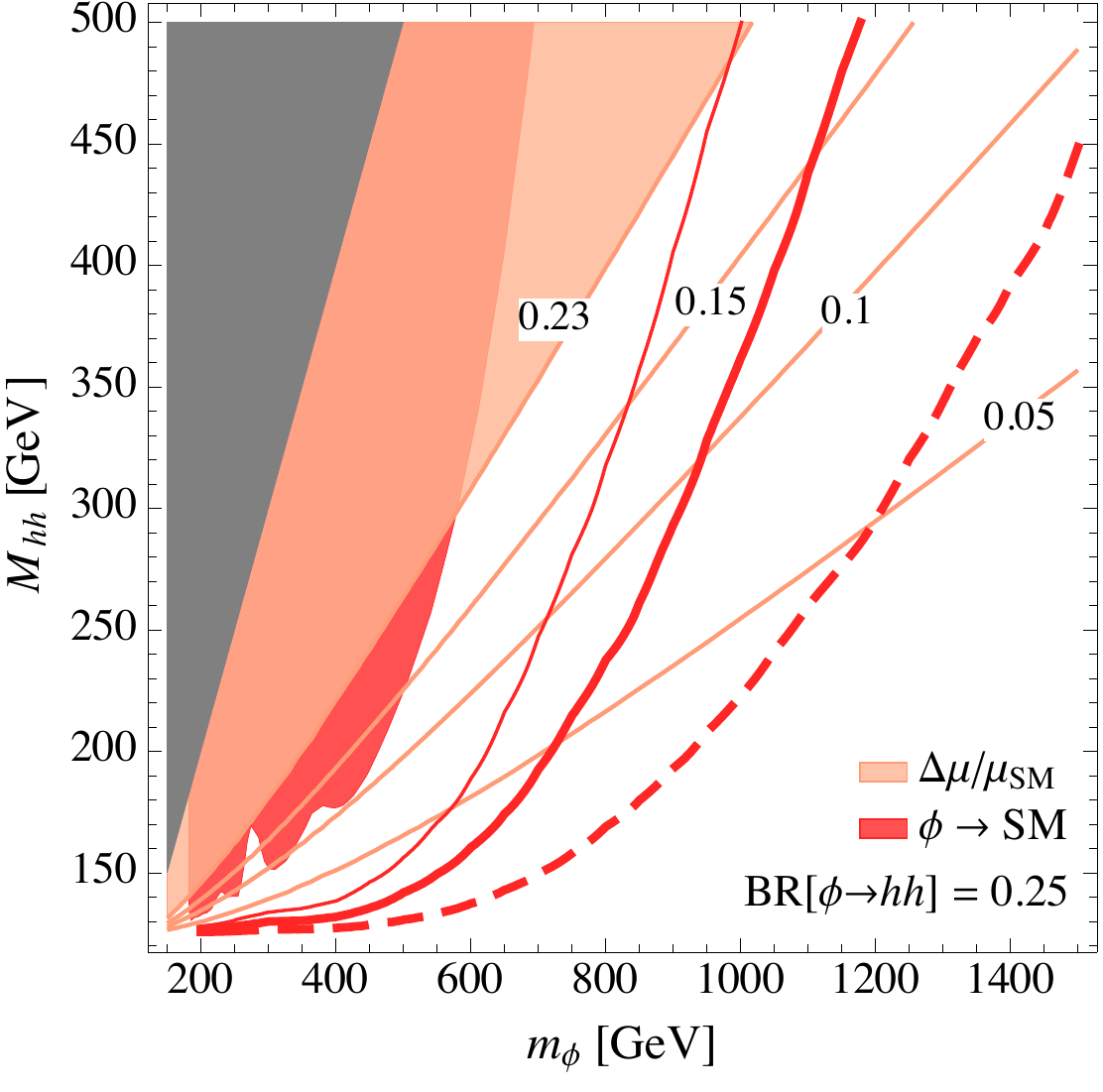}\hfill%
\includegraphics[width=.49\textwidth]{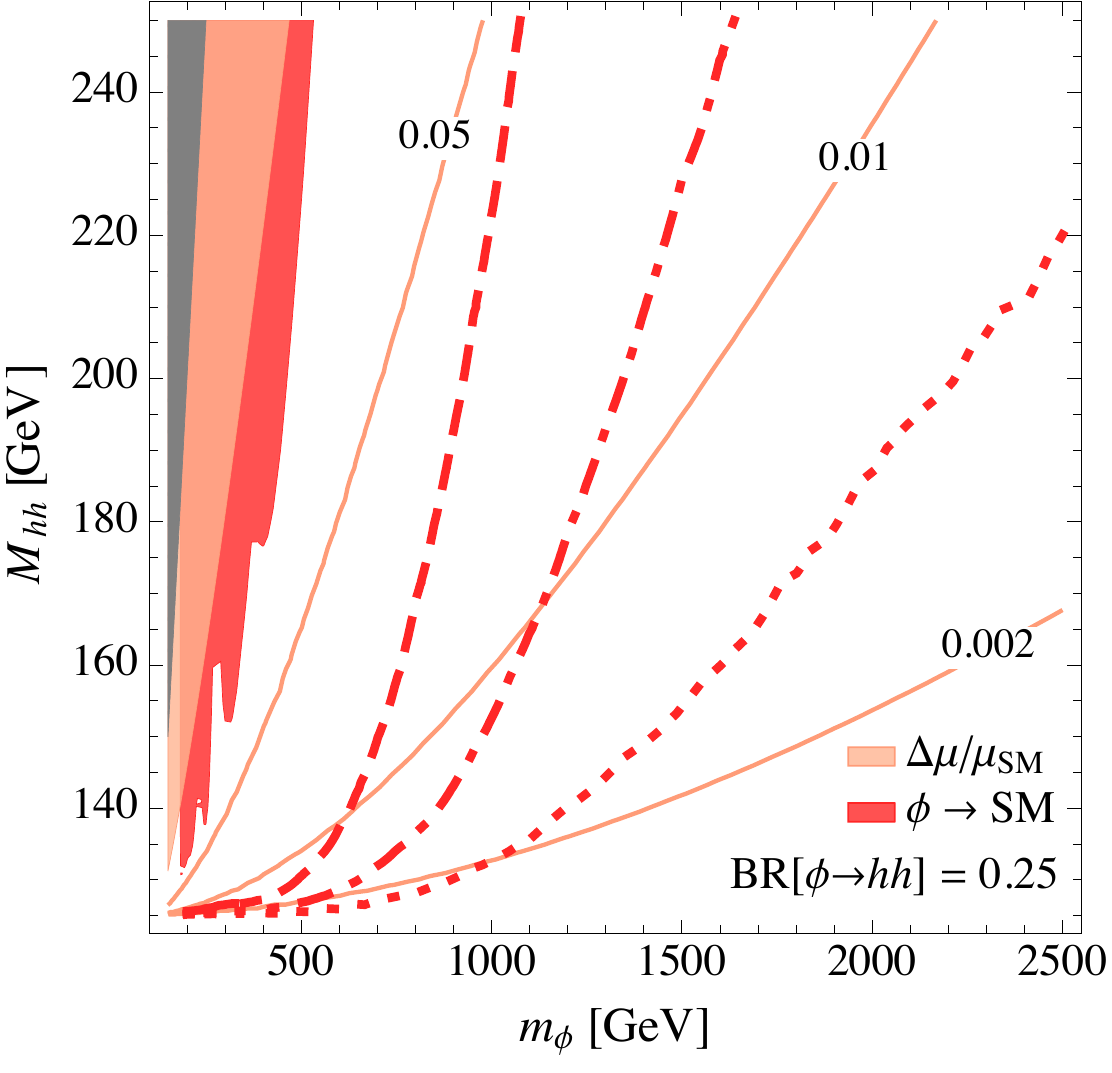}
\caption{Left: Comparison between the combined direct LHC exclusion and reaches (red) and the deviations in Higgs couplings (pink), with BR$_{\phi\to hh}$ = 0.25. Right: the same comparison, in the region relevant for future colliders. The coloured regions are excluded at 95\% C.L.\ and the notation for the lines is the same as in \figuresnames s~\ref{fig:direct75} and \ref{fig:direct250}.\label{fig:direct_indirect}}
\end{figure}

In \figuresnames~\ref{fig:direct_indirect} we show a comparison of all the direct searches with the indirect constraints from the measurement of Higgs signal strengths. Here we fix ${\rm BR}_{\phi\to hh}$ to the asymptotic value of 0.25, and the direct exclusion is always dominated by $\phi\to VV$. As in \figuresnames s~\ref{fig:triple1} and \ref{fig:triple2}, we draw contours of constant $s_\gamma^2$ instead of explicit indirect exclusions for each future collider, since the projections for the precisions that can be attained on the Higgs couplings are still quite uncertain. One can however easily read the numbers off table~\ref{tab:Higgs_couplings}. Notice that in the right-handed plot the range of the vertical axis is changed in order to better show the region of small mixing angle most relevant for future colliders.

The direct searches are always more powerful than the Higgs signal strength measurements for low values of the resonance mass, while the opposite is true for high masses. While this is a general feature of direct and indirect searches, such an interplay was a priori not guaranteed in the physically interesting range of masses, and for the actual sensitivities of the future experiments. At the LHC, the point where the two searches become comparable varies from $m_\phi \sim 500$ GeV for LHC8 to slightly more than 1 TeV at HL-LHC. Also a linear $e^+e^-$ collider, with a sensitivity to deviations in the Higgs couplings in the 1\% ballpark, would become more sensitive than the HE-LHC upgrade above about 1 TeV (500 GeV for HL-LHC). The same mass of a TeV sets also the limit between FCC-ee and FCC-hh.

\section{Non minimal supersymmetry}\label{sec:susy}
We can now apply the previous general discussions to some concrete models that are particularly motivated scenarios for physics beyond the SM.
In this section we consider the Next-to-Minimal Supersymmetric Standard Model (NMSSM), which is the minimal deformation of the MSSM that includes a scalar singlet in the spectrum. A chiral superfield $S$, singlet under the SM gauge groups, is coupled to the two Higgs doublets $H_u$ and $H_d$ of the MSSM through the superpotential
\begin{equation}
\label{superpotential}
\W_{\rm NMSSM} = \W_{\rm MSSM} + \lambda S H_u H_d + f(S),
\end{equation}
where $f(S)$ is a generic polynomial up to third order in $S$.

This deformation has important consequences for the naturalness of the theory, given the measured value of the Higgs mass. Indeed, in the MSSM a 125 GeV Higgs requires the largest contribution to $m_h$ to originate from SUSY-breaking effects, $m_h^2 \lesssim m_Z^2 +\Delta^2$, where $\Delta^2$ is the radiative correction mainly due to the top-stop sector (see {\it e.g.}\ \cite{Carena:1995wu}). This in turn requires either top squarks in the multi-TeV range, or large trilinear terms, both cases implying a large tuning of the electroweak scale. The ultimate reason for this is that the supersymmetric couplings in the MSSM are controlled by the weak gauge couplings. The situation is different in the NMSSM, where a supersymmetric contribution to the Higgs mass originates at tree-level from the Yukawa-like term $\lambda S H_u H_d$, 
\begin{equation}\label{mh}
m_h^2 \lesssim \frac{\lambda^2 v^2}{2} s_{2\beta}^2 + m_Z^2 c_{2\beta}^2 +\Delta^2,
\end{equation}
where $\tan\beta \equiv v_u/v_d$ is the ratio of the vev's of the two Higgs doublets $H_{u,d}$. As manifest from \eqref{mh}, the most natural scenario occurs when the coupling $\lambda$ is of order one, so that the largest contribution to $m_h$ has a supersymmetric origin. We here briefly sketch the major features of such a picture:

\begin{itemize}
\item[$\diamond$] It gives a large supersymmetric contribution to the Higgs mass (with a different dependence on $\tan\beta$ than the D-terms) that is sufficient to achieve 125 GeV at tree-level. 
\item[$\diamond$] It modifies the dependence of the electro-weak vacuum expectation value on the Lagrangian parameters, allowing for parametrically larger soft masses of stops and gluinos  -- by a factor of $\O(2\lambda/g)$ -- than in the MSSM, for a given amount of tuning \cite{Hall:2011aa,Gherghetta:2012gb,Agashe:2012zq}.
While a large $\lambda > 1$ softens the tuning in $v$, it increases the tuning in the Higgs mass because it overshoots its measured value.
\item[$\diamond$] The coupling is not asymptotically free and if $\lambda\gtrsim 0.7$ the theory undergoes a strong coupling regime at a scale lower than the GUT scale \cite{Espinosa:1991gr}. We will consider both regimes in the following, as they both have been deeply studied in the literature.
\end{itemize}
The NMSSM with a sizeable $\lambda$ -- which is the leading parameter also for phenomenology -- opens then the interesting possibility that the lightest new particles be the extra scalar bosons of its extended Higgs sector (possibly with the exception of the LSP). 
These new physical degrees of freedom consist in three neutral scalars, two pseudoscalars, and one charged Higgs boson.

The phenomenology of the NMSSM is controlled by several parameters, so that its study is often performed via scans on some motivated ranges (see {\it e.g.}\ \cite{King:2014xwa}), which need the specification of a particular version of the NMSSM potential, {\it i.e.}\ the function $f(S)$ and all the soft terms. Here we adopt a more general approach to have an analytic understanding of the phenomenological properties of the extra Higgs bosons, focussing on the CP-even ones, in the spirit of \cite{Barbieri:2013hxa,Barbieri:2013nka}.

\begin{figure}[t]
\centering%
\includegraphics[width=0.49\textwidth]{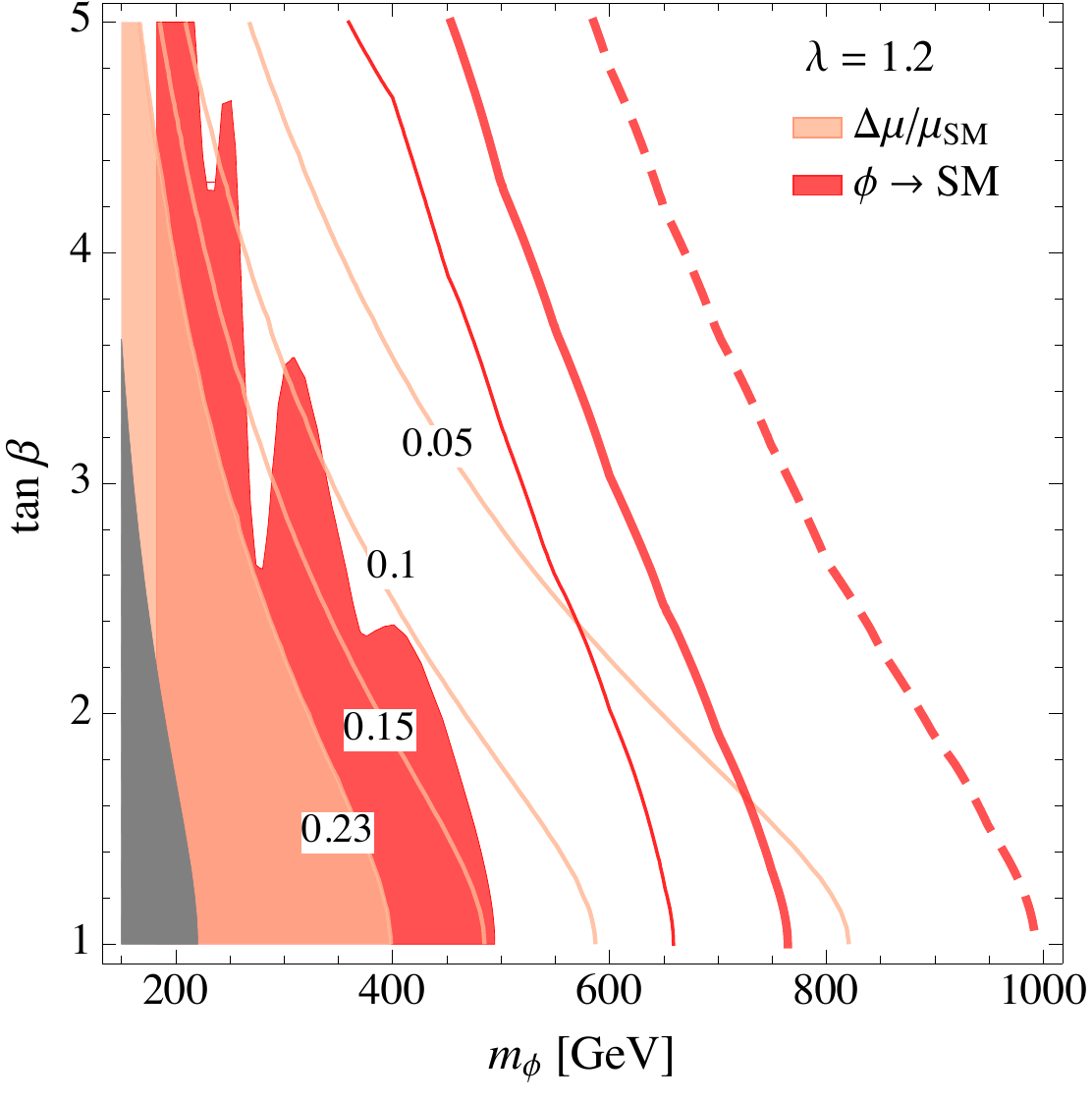}\hfill%
\includegraphics[width=0.49\textwidth]{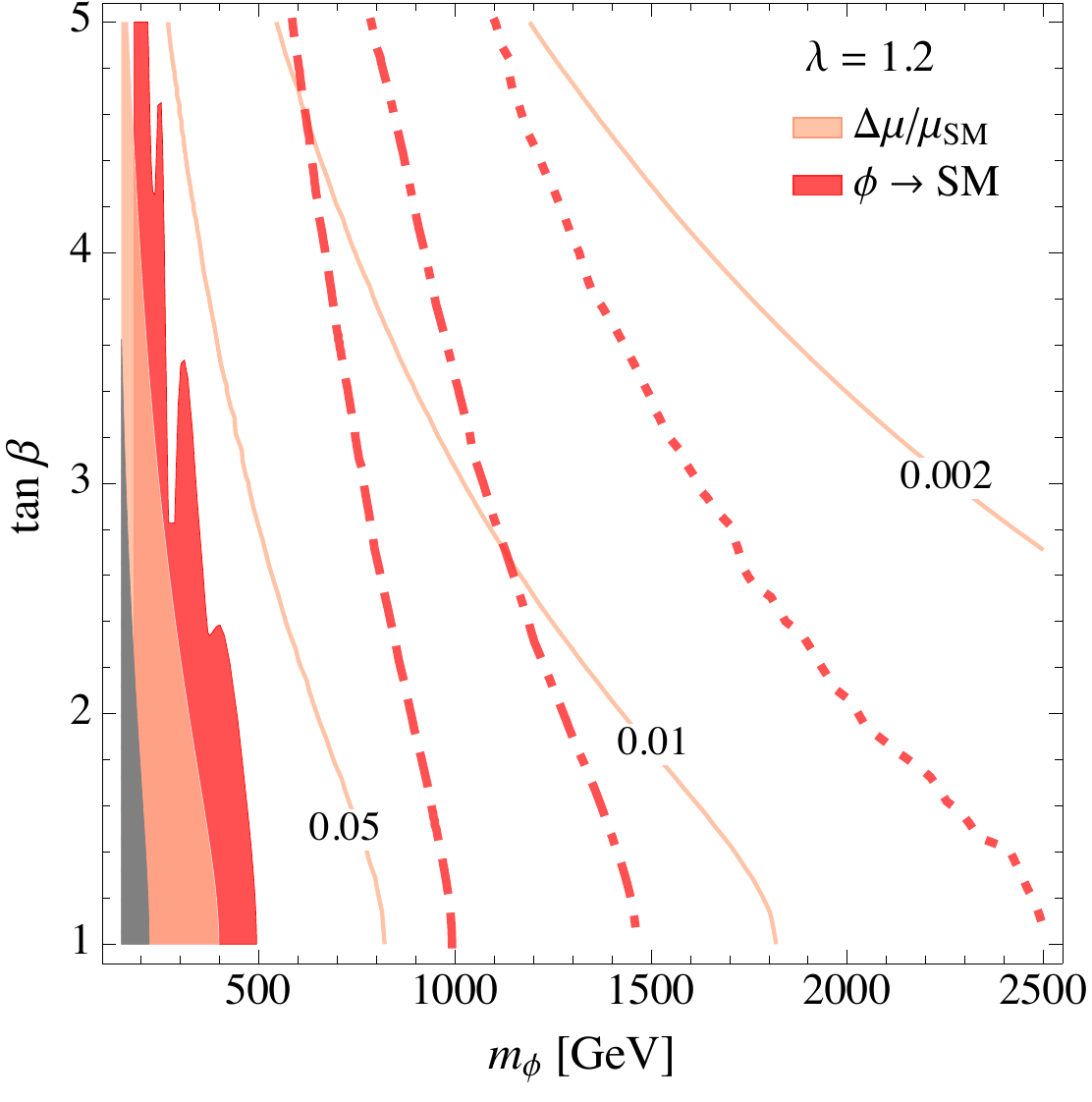}
\caption{\label{fig:NMSSM_1} NMSSM with ``strong'' coupling, $\lambda=1.2$ and $\Delta=70$ GeV. Shaded regions and lines as in \figuresnames~\ref{fig:direct_indirect}.}
\end{figure}

The SM-like Higgs field of \eqref{fields} reads $H = \sin\beta\,H_u + \cos\beta\,H_d$. As the present data suggest (see \cite{Barbieri:2013hxa,Barbieri:2013nka,Barbieri:2013aza}), a particularly interesting limit is the one where the orthogonal Higgs doublet, which does not participate in EWSB, is decoupled.\footnote{The case of a non-decoupled second doublet is mostly probed via Higgs coupling measurements, and its phenomenology is similar to that of an MSSM Higgs\cite{Barbieri:2013hxa}. Moreover, such possibility is further constrained by the presence of a charged Higgs boson.} 
Under the assumption of CP conservation, we can single out the CP-even component from the complex field $S$. In this case the system is equal to the two-state scenario described in \sectionname~\ref{sec:Singlet_generic}, and can be studied along the same lines.

The main parameters are again $M_{hh}^2$ and $m_\phi$, with the former depending on the parameters of the model like
\begin{equation}
M_{hh}^2= \frac{\lambda^2v^2}{2}\sin^2 2\beta + m_Z^2\cos^2 2\beta  + \Delta^2.
\label{Mhh_NMSSM}
\end{equation}
Similarly to \sectionname~\ref{sec:Singlet_generic}, the mass eigenstates $(h,\phi)$ are defined as in \eqref{rotation}, while the mixing angle $\gamma$ between the Higgs and the singlet is now given by
\begin{equation}\label{singammaNMSSM}
\sin^2 \gamma = \frac{m_Z^2\cos^2 2\beta + (\lambda^2v^2/2)\sin^2 2\beta + \Delta^2 - m_h^2}{m_\phi^2 - m_h^2}.
\end{equation}
The phenomenology is mainly set by $\gamma$ and $m_\phi$ but, as discussed in \sectionname~\ref{sec:parametrization}, quantities like the trilinear couplings depend on other parameters. For simplicity, in the following we will fix BR$_{\phi \to hh}$ to its asymptotic value 1/4, so that there is no need to enter the details of the scalar potential.\footnote{\label{foot:NMSSM_potential}The coefficients of the potential in \eqref{potential} can be computed in terms of the parameters of the model. This amounts to have an explicit expression for $f(S)$ in \eqref{superpotential} and for the soft term lagrangian $\mathcal{L}_{\rm soft}$. As an example in the scale invariant NMSSM (see \cite{Ellwanger:2009dp} for a review),
\[ \mathcal{W}= \lambda S H_u H_d + \frac{\kappa}{3}S^3,\quad\quad -\mathcal{L}_{\rm soft} \supset A_\lambda (S H_u H_d + h.c.) + A_\kappa (S^3 + {\rm h.c.})  \]
The matching for the coefficients, upon the decoupling of one of the two Higgs doublets, is given by
\[ \lambda_{H}= \frac{\lambda^2}{4} \sin^2(2\beta) + \frac{m_Z^2}{2v^2}\cos^2(2\beta) + \frac{\Delta^2}{2v^2},\quad \lambda_{S}= |\kappa|^2,\quad \lambda_{HS}= |\lambda|^2,\quad a_H=A_\lambda \sin(2\beta), \quad a_S = A_\kappa .\]}
Notice however that all the analysis of \sectionname~\ref{sec:potential} applies, in general, also to the present case. In particular, from the point of view of phenomenology $v_s$ is again the leading extra parameter.


\subsection{Results and discussion}

\begin{figure}[t]
\centering%
\includegraphics[width=0.49\textwidth]{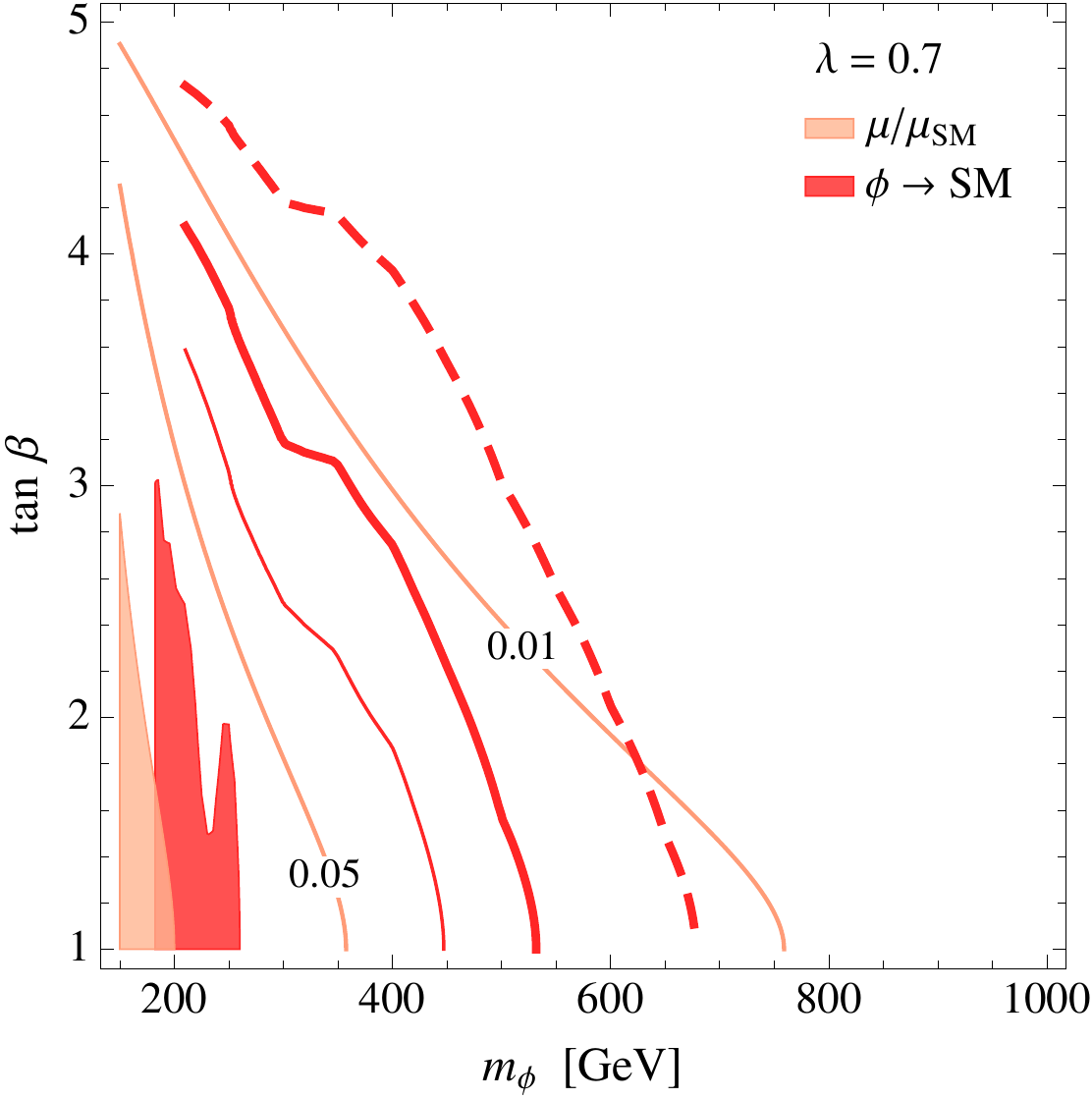}\hfill%
\includegraphics[width=0.49\textwidth]{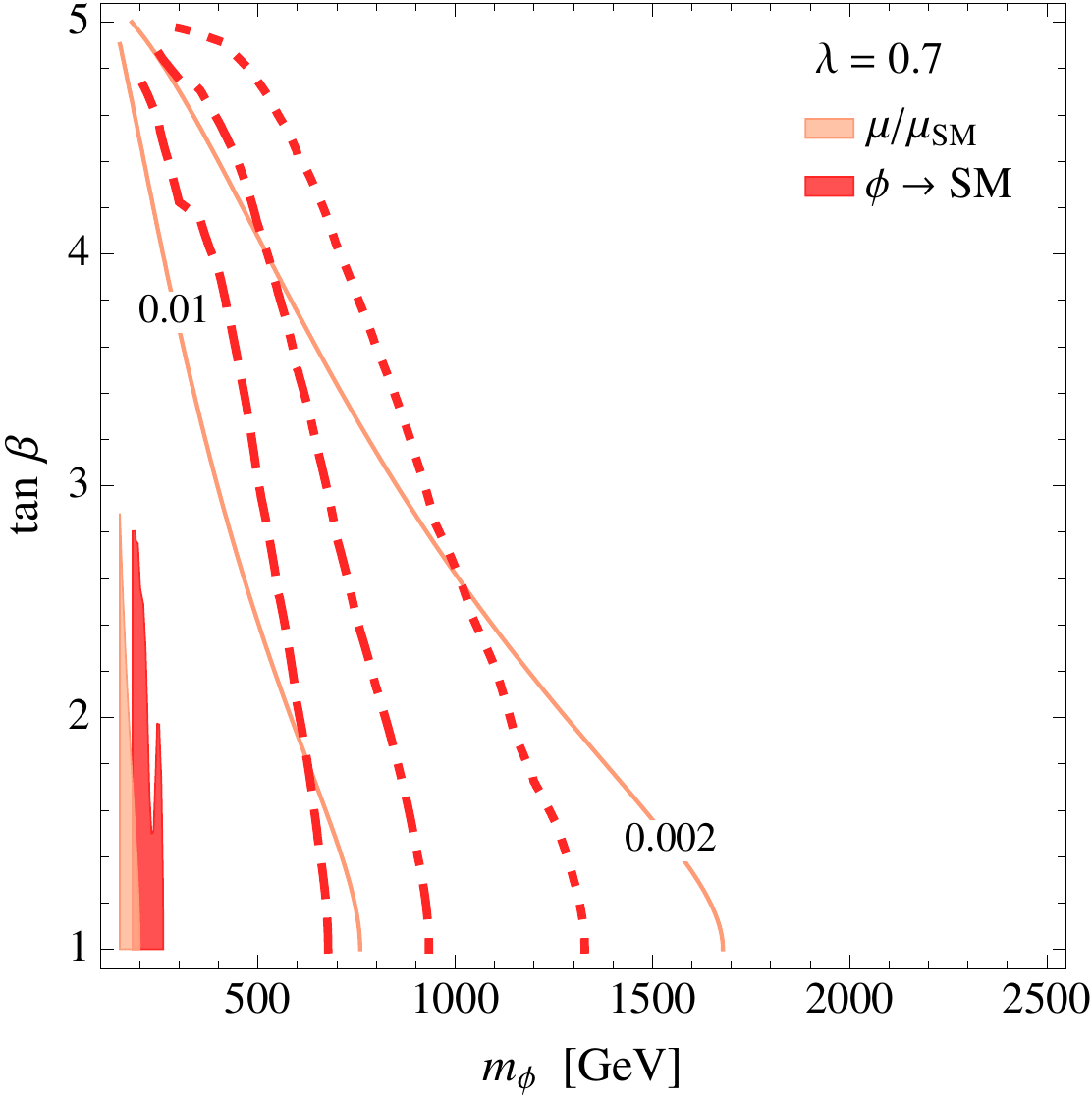}
\caption{\label{fig:NMSSM_2} NMSSM with ``perturbative'' coupling, $\lambda=0.7$ and $\Delta=80$ GeV. Shaded regions and lines as in \figuresnames~\ref{fig:direct_indirect}.}
\end{figure}

The quantity $M^2_{hh}$ now depends on parameters with a defined physical interpretation.
Given the expression \eqref{singammaNMSSM} we have chosen to fix $(\lambda,\Delta)$ and let $\tan\beta$ free to vary. Notice that the value of $\Delta$ is almost irrelevant for sizeable $\lambda$. We consider the following two cases:
\begin{itemize}
\item[$\diamond$] ``Strong'' coupling, $\lambda=1.2$.
This scenario is realised for example in $\lambda$-SUSY \cite{Barbieri:2006bg}. The contribution from radiative corrections has been fixed to $\Delta=70$ GeV. This choice  is consistent with a mass of the lightest stop both at the edge of the present exclusion \cite{ATLAS-CONF-2015-010,Khachatryan:2015wza}, for maximal mixing, and in the multi-TeV range.
\item[$\diamond$] Perturbative coupling, $\lambda=0.7$.
In this case we use a value for $\lambda$ consistent with perturbative unification at the GUT scale, and we take $\Delta= 80$ GeV in order to more easily reproduce the Higgs mass. 
\end{itemize}

Figures~\ref{fig:NMSSM_1} and \ref{fig:NMSSM_2} show the present and future constraints on this scenario for both indirect and direct searches. We restrict ourselves to low values of $\tan\beta$ since this is the region most relevant for naturalness \cite{Hall:2011aa,Gherghetta:2012gb}.
Contrary to the generic Singlet case, the direct searches always dominate at the LHC and its upgrades. In the context of \figuresnames~\ref{fig:direct_indirect}, this can be interpreted with the observation that, for the physically motivated ranges of NMSSM parameters, $M_{hh}$ cannot be too large, see \eqref{Mhh_NMSSM}. The two benchmarks with different values of $\lambda$ have significantly different behaviours: the larger is $\lambda$ the smaller is the allowed parameter space.  Moreover, while for a larger $\lambda$ leptonic colliders will start to be competitive with hadronic ones, for a smaller $\lambda$ only a per-mille level of precision in the Higgs couplings could be competitive with direct searches. This means that the observation of a deviation in the Higgs signal strengths, without any direct indication of an extra scalar, will strongly disfavour this scenario.
It is interesting that, for supersymmetry at the LHC, our findings indicate that the strategies to best look for a singlet-like Higgs and a doublet-like one look quite different: direct searches in the first case, and mainly indirect ones in the second case (barring the large $\tan\beta$ case \cite{Khachatryan:2014wca,Aad:2014vgg}).

From the point of view of indirect searches, it is notable that also here large deviations from the SM value are possible in the triple Higgs coupling, as shown in \cite{Barbieri:2013hxa} and confirmed in a particular model by \cite{Cao:2014kya}, and contrary for example to the MSSM case. 
In general there are no particular differences with respect to the analysis of \sectionname~\ref{sec:Singlet_generic}, given that $v_s$, $\lambda$ and $\lambda_S$ are free parameters also in the NMSSM.
However, as we have already discussed, the NMSSM prefers low values of $M_{hh}$.
Accordingly, the deviations from one in $g_{hhh}/g_{hhh}^{\rm SM}$ are expected to be milder than in the generic case (see figures~\ref{fig:triple1} and \ref{fig:triple2}), but still allowing for observable effects at the LHC.


\section{Twin \& Composite Higgs}\label{sec:twin}
The presence of an additional scalar singlet is also typical of another class of models, namely those where the Higgs is a pseudo-Goldstone boson. In this type of models the Higgs boson emerges as an approximate Goldstone boson associated to the breaking of some global symmetry $\mathcal{G} \to \mathcal{H}$. In the case where there is only one physical pseudo Goldstone Higgs, the only other additional scalar singlet is the ``radial mode'', {\it i.e.}\ the degree of freedom associated with vacuum expectation value $f$, with $f > v$, that breaks the global symmetry. Throughout this section we dub the extra physical scalar $\sigma$ instead of $\phi$.

If the above description derives from a weakly coupled model at higher energies, then the radial mode $\sigma$ could be relatively light, significantly below $4\pi f$. This is for example the case of the Twin Higgs (TH) idea \cite{Chacko:2005pe}, with a weakly coupled description \cite{Barbieri:2005ri, Falkowski:2006qq,Chang:2006ra,Craig:2013fga,Craig:2015pha}, and we now turn to this possibility.

\paragraph{Twin Higgs.}
Twin Higgs models solve the little hierarchy problem up to some scale $\Lambda$, without requiring the presence of coloured degrees of freedom below such scale.\footnote{Above $\Lambda$, a theory solving the big hierarchy between $\Lambda$  and the Planck (or GUT) scale is needed.} This is contrary to the usual realisations of SUSY and composite Higgs models (CHM), where new coloured bosons or fermions are expected to lie within the LHC reach. In TH models the low energy degrees of freedom that cancel the ``quadratic divergences'' are total singlets of the SM and hence very difficult to detect at the LHC, while coloured particles are expected to be naturally heavier. In TH models even a null result from the LHC will leave the tuning of the weak scale at the level of $\O(10\%)$ (for recent studies see \cite{Craig:2015pha,Burdman:2014zta,Geller:2014kta,Barbieri:2015lqa,Low:2015nqa}). 

In this paper we would like to emphasise that, if the TH has a weakly coupled description, an interesting possibility consists in trying to detect the ``radial mode'', the mirror partner of the Higgs. In fact, differently from other mirror particles, it can be singly produced through a mixing with the Higgs. This is complementary to the idea that the only model-independent probe of such scenarios consists of the Higgs signal strengths. In order to discuss the phenomenology of this picture we here briefly review the basics of the model.
\begin{itemize}
\item[$\diamond$] A copy of the SM gauge, matter and scalar content is added, related to the SM by a $Z_2$ symmetry. A scalar portal coupling is allowed between the two sectors.
\item[$\diamond$] An important assumption is that the scalar potential is accidentally SO(8) invariant,\footnote{For our discussion also SU(4)/SU(3) is fine.} in such a way we can have a spontaneous symmetry breaking SO(8)/SO(7) parametrized by\\
\begin{equation}
V_{\rm symm}(H_A,H_B) = \lambda_* \Big( |H_A|^2 + |H_B|^2 - \frac{f_0^2}{2} \Big)^2\,,
\label{eq:V_GtoH}
\end{equation}
where $H_A$ is the SM Higgs doublet, and $H_B$ is its $Z_2$-symmetric partner, singlet under the SM gauge group. The global symmetry SO(8) is manifest when the above expression is written in terms of the real components. From \eqref{eq:V_GtoH}, and from the fact that the two copies of the SM are gauged, out of the 7 Goldstones only one remains as a physical fluctuation, the other six being ``eaten'' by the vector fields. In addition there is also a radial mode of mass $\sim \sqrt{\lambda_*} f_0$. At this level one would like to explicitly break both SO(8), to give a mass also to the other degree of freedom, and $Z_2$, in order to get a hierarchy between the two vev's in the two sectors.
\item[$\diamond$] This can be achieved by adding to the above potential other gauge-invariant terms, so that the full tree-level potential $V$ reads
\begin{equation}
V(H_A,H_B) =  \kappa \big(|H_A|^4 + |H_B|^4\big) + m^2 (|H_A|^2 - |H_B|^2) + V_{\rm symm}(H_A,H_B)\,,
\label{eq:V_TH_tot}
\end{equation}
where the first term breaks SO(8) but preserves $Z_2$, and the second one breaks both groups. A nice feature of TH models is that the radiative corrections do not introduce any sensitivity to high scales, and the finite contributions are suppressed by the fourth power
of SM couplings (as an example $\kappa\sim \O(g_{\rm SM}^4)$ \cite{Chacko:2005pe}).\footnote{\label{foot:potential}A term $\kappa^\prime |H_A|^2 |H_B|^2$ can be reabsorbed in a redefinition of $\kappa$ and of the coefficients in \eqref{eq:V_GtoH}. The same is true for quadratic terms containing a different combination of $|H_A|^2$ and $|H_B|^2$.}
\end{itemize}
These three ingredients are sufficient to describe the phenomenology. The parameters in \eqref{eq:V_TH_tot} are related to the vev's $\langle H_A^2 \rangle \equiv v^2/2$ and $\langle H_B^2 \rangle \equiv (f^2 - v^2)/2$ via
\begin{equation}\label{eq:VEVs}
v^2 = \frac{\kappa  \lambda_*\,f_0^2 - (\kappa+2  \lambda_*)\,m^2}{\kappa(\kappa + 2 \lambda_*)},\quad f^2 = f_0^2 \frac{2 \lambda_*}{2\lambda_*+\kappa},
\end{equation}
so that $f$ is the symmetry breaking scale of a strongly-coupled scenario, which indeed is obtained in the limit of large $\lambda_*$, {\it i.e.}\ when the mass of the radial mode is close to the cutoff and it can be consistently integrated out.

From the diagonalisation of the mass matrix we can then fix $m_h$ and trade the residual free parameter for $m_\sigma$, the mass eigenvalue of the singlet-like state.
The mass eigenstates $h$ and $\sigma$ are related to the gauge ones via a rotation by an angle $\gamma$ as in \eqref{rotation}, with
\begin{equation}\label{singammaTWIN}
\sin^2\gamma= \frac{v^2}{f^2}- \frac{m_h^2}{m_\sigma^2 -m_h^2}\Big(1- 2 \frac{v^2}{f^2}\Big).
\end{equation}
In the notation of \sectionname~\ref{sec:Singlet_generic}, the parameter $M_{hh}^2$ is now given by
\begin{equation}
M^2_{hh}=\frac{v^2}{f^2}(m_\sigma^2 + m_h^2).
\end{equation}
A crucial observation is that the only two free parameters are ($m_\sigma$, $f$), which completely determine the phenomenology.
Like in the generic singlet case of \sectionname~\ref{sec:Singlet_generic}, the mixing angle is also the value of the normalised couplings of $h$ and $\sigma$ to SM particles. Other important informations are provided by the trilinear couplings of the mass eigenstates
\begin{equation}
\begin{split}
\frac{g_{hhh}}{g_{hhh}^{\rm SM}}&= \frac{1-2 v^2/f^2}{\sqrt{1-v^2/f^2}}\bigg[1+\frac{3 m_h^2}{2 m_\sigma^2} + \O\Big(\frac{m_h^4}{m_\sigma^4}\Big)\bigg],\\
g_{\sigma hh}&=\frac{m_\sigma^2}{f} \bigg[ 1-\frac{m_h^2 \left(f^4-8 f^2 v^2+8 v^4\right)}{2 m_\sigma^2\, v^2 \left(f^2-v^2\right)}+ \O\Big(\frac{m_h^4}{m_\sigma^4}\Big)\bigg].
\end{split}
\label{TH_Higgs_couplings}
\end{equation}


\paragraph{Composite Higgs models.}
The above model only describes two real scalars, so that it closely resembles the case of a linearised minimal composite Higgs model with an SO(5)/SO(4) symmetry breaking \cite{Agashe:2004rs}. In this context the global symmetry is spontaneously broken by some strong interaction, and the Higgs is again a pseudo-Goldsone boson of this breaking. The main phenomenological difference with the previous case is the presence of vector and fermion resonances related with the strong sector, with a mass of a few times the scale $f$.\footnote{See {\it e.g.}\ \cite{Contino:2010rs} and references therein.}

There is one single radial mode $S$ associated with this symmetry breaking pattern, which forms a quintuplet of SO(5) together with the Higgs doublet $H$; at energies higher than its mass one can describe this scenario through a linear $\sigma$-model.
Although this might be an improper description of a strongly-interacting scenario, where the radial mode is generically expected to have a mass close to the cut-off of the theory $4\pi f$, it provides a useful and simple parametrisation \cite{Barbieri:2007bh}. Using the same notation as above, and trading $H_B$ with the pure real singlet $S$, the potential can be written as \cite{Contino:2011np,Barbieri:2013aza}
\begin{equation}\label{eq:V_composite}
V(H,S) = \lambda_* (|H|^2 +S^2 - f_0^2)^2 +\alpha f_0^2 |H|^2 -\beta |H|^2 S^2,
\end{equation}
and it is not difficult to show that, to describe the scalar degrees of freedom, the potential in (\ref{eq:V_composite}) is equivalent to the TH one in (\ref{eq:V_TH_tot}) (see {\it e.g.}\ footnote \ref{foot:potential}).

\begin{figure}[t]
\begin{center}
\includegraphics[width=.49\textwidth]{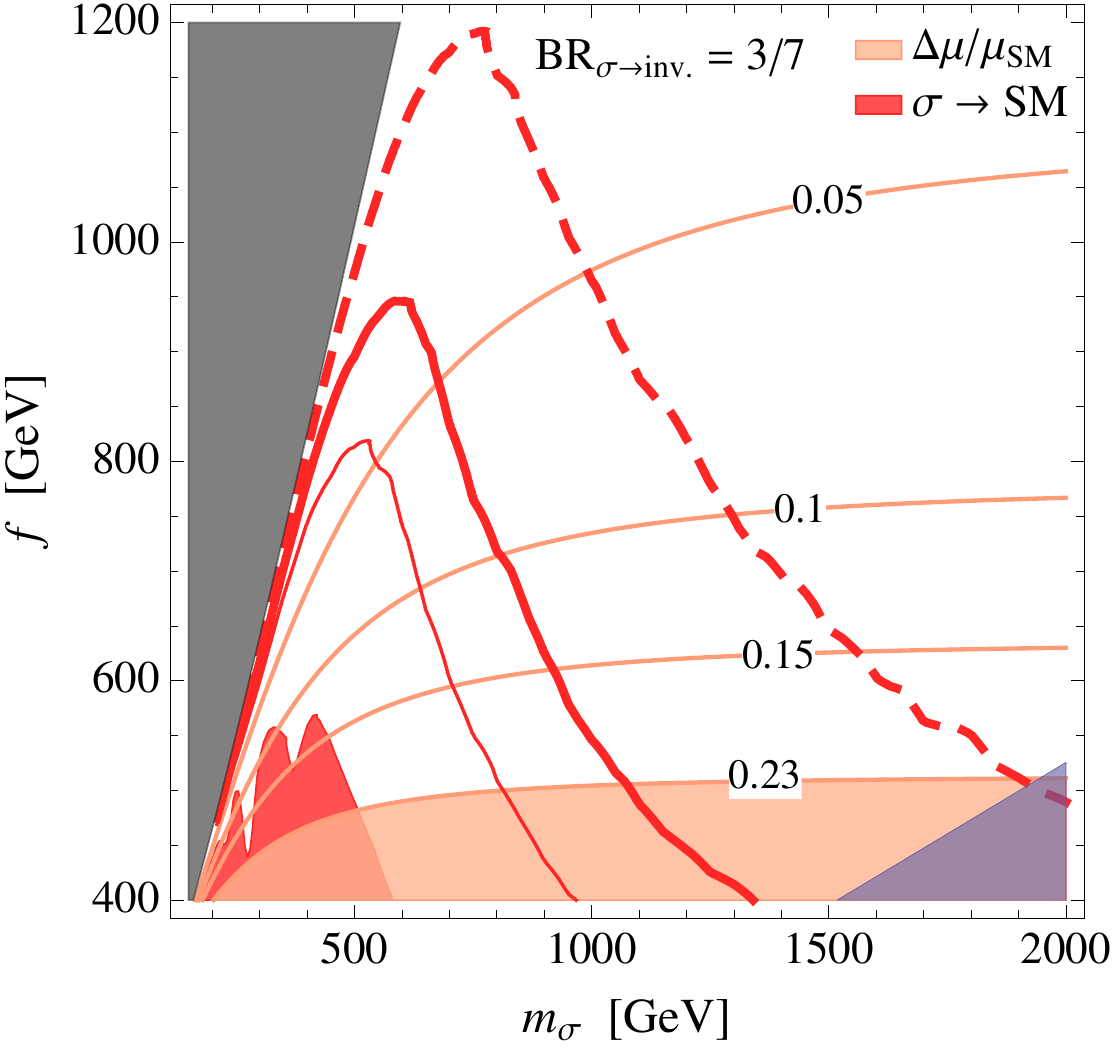}\hfill%
\includegraphics[width=.49\textwidth]{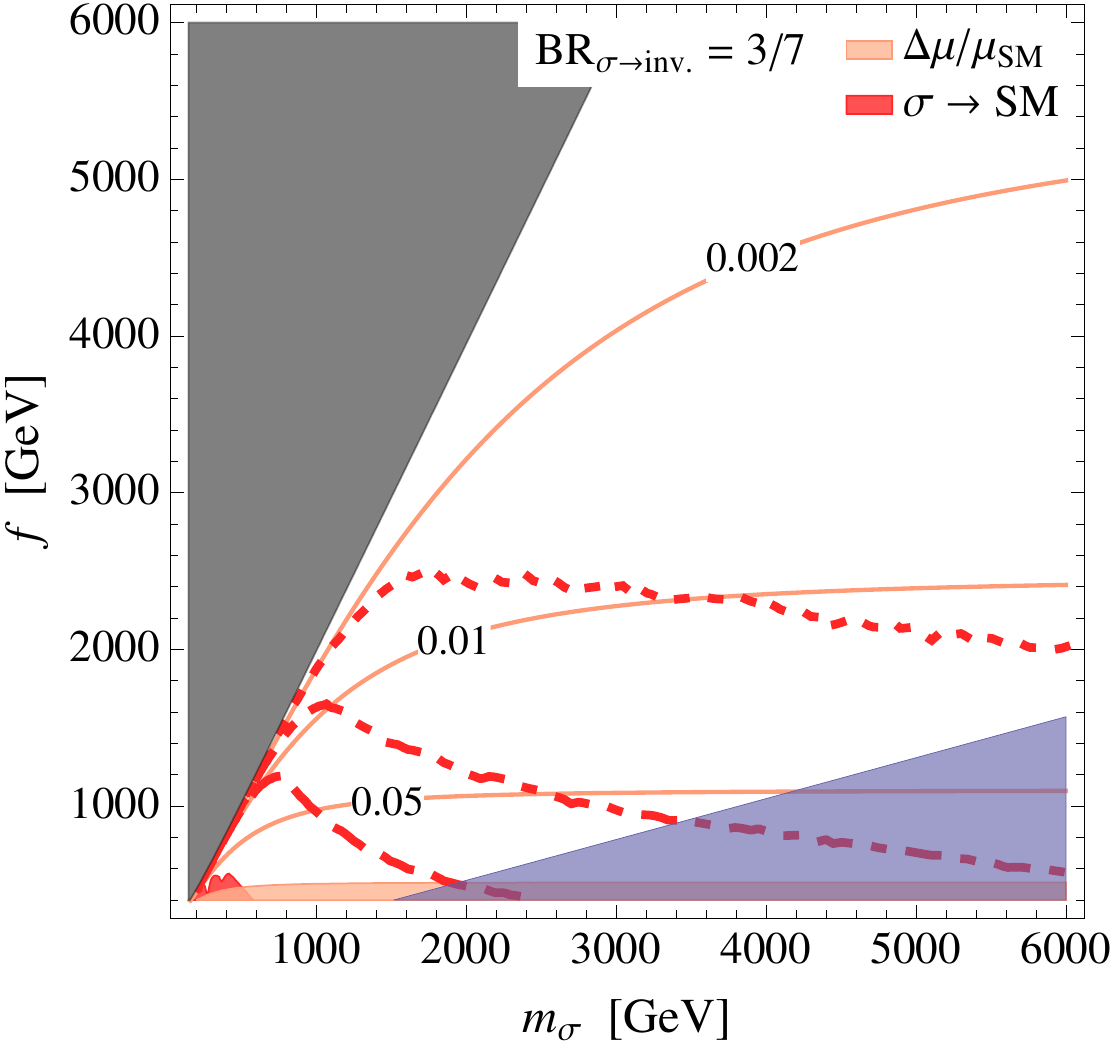}
\caption{\label{fig:twin}\small Model with $\mathrm{BR}_{\sigma \to{\rm inv.}}=3/7$. Shaded regions: excluded at 95\% C.L. by Higgs couplings (pink), excluded by direct searches (red), $\Gamma_\sigma > m_\sigma$ (blue), unphysical parameters (grey). The notation for the lines is as in \figuresnames~\ref{fig:direct_indirect}.}
\end{center}
\end{figure}


\subsection{Results and discussion}
\label{sec:twinpheno}

\begin{figure}[t]
\begin{center}
\includegraphics[width=.49\textwidth]{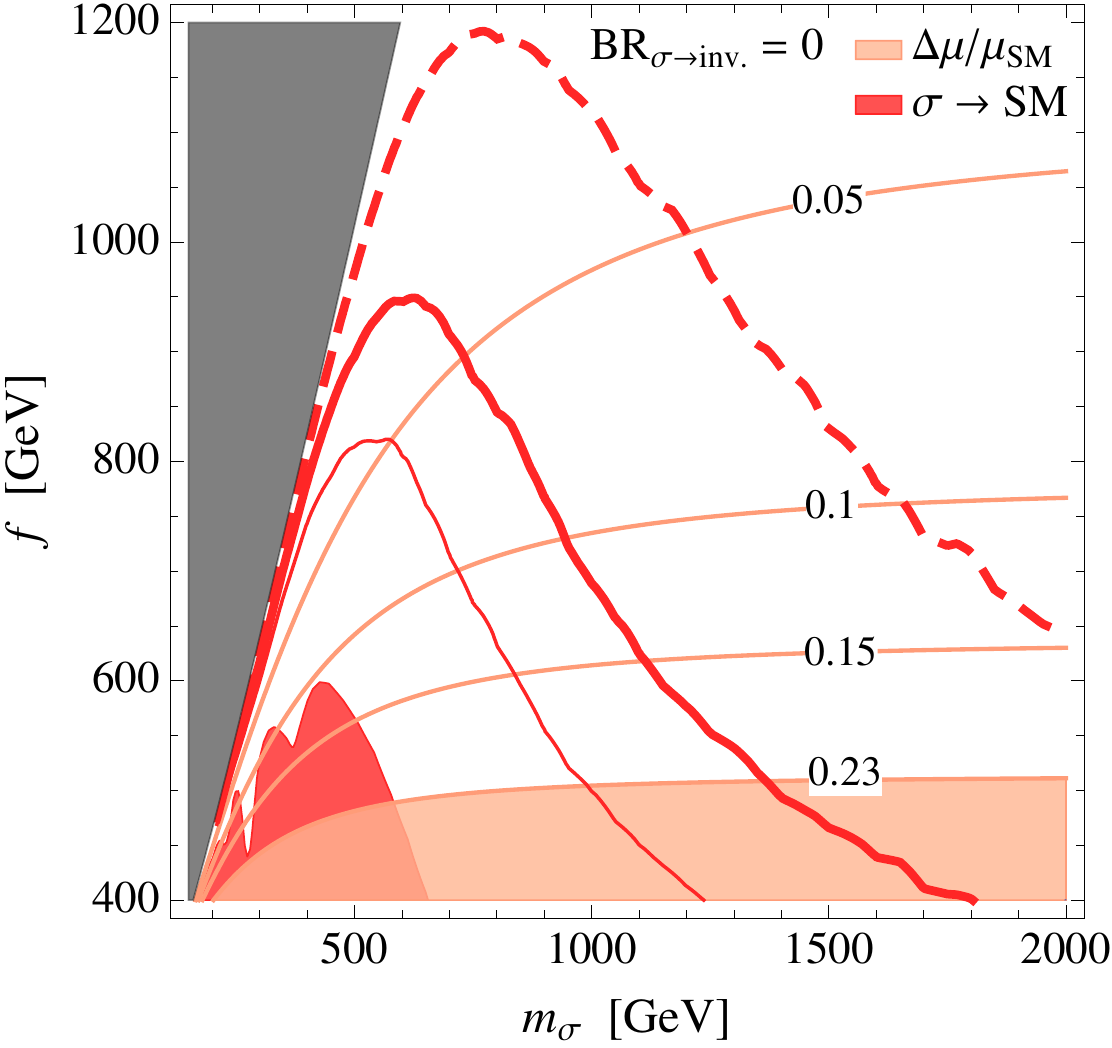}\hfill%
\includegraphics[width=.49\textwidth]{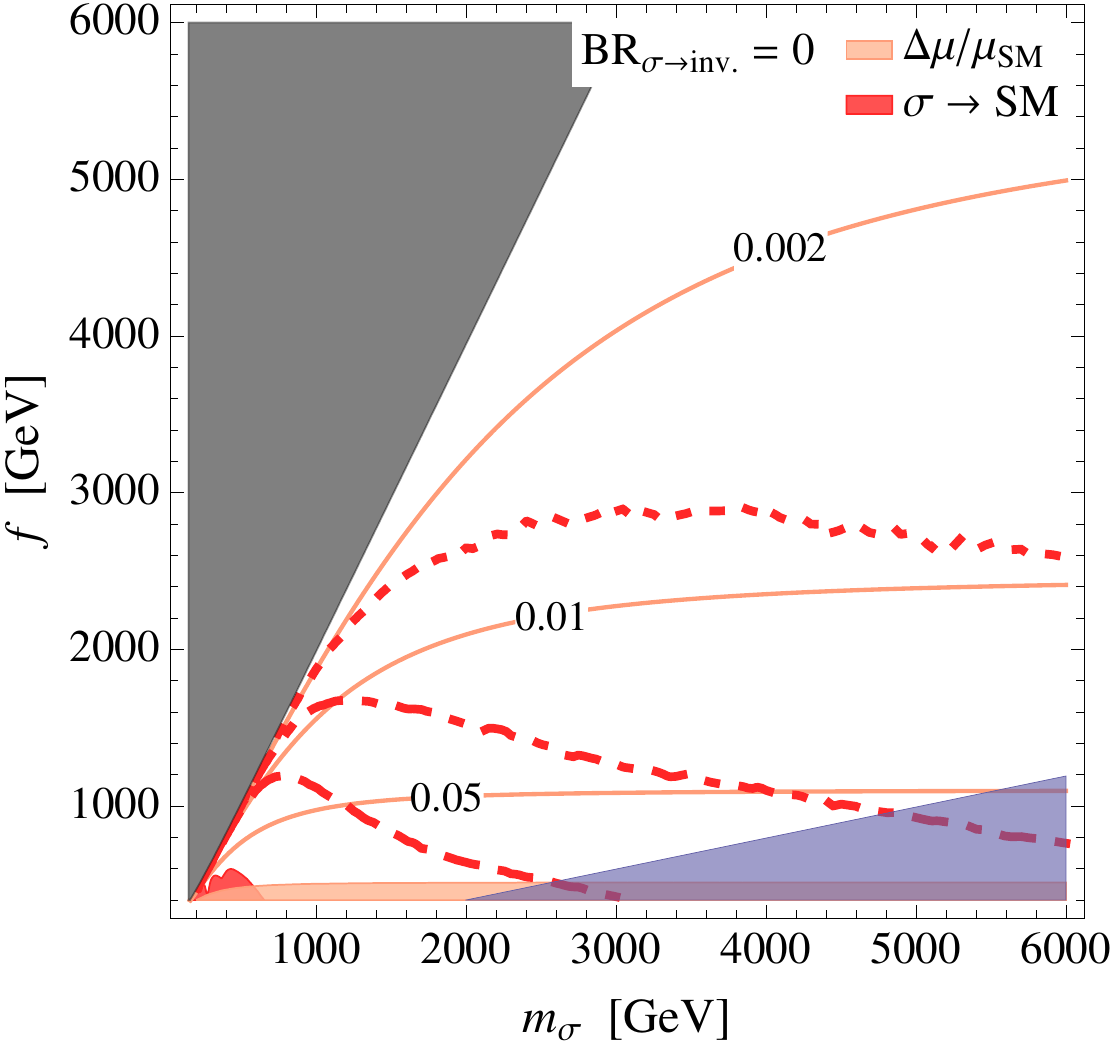}
\caption{\label{fig:chm}\small Model with $\mathrm{BR}_{\sigma \to{\rm inv.}}=0$. Notation as in \figuresnames~\ref{fig:twin}.}
\end{center}
\end{figure}

Given the discussion above, the following phenomenological analysis will apply to the scalar sectors of both Twin (\ref{eq:V_TH_tot}) and composite \eqref{eq:V_composite} Higgs models. There are two main differences:
\begin{itemize}
\item[$\circ$] in the case of the ``composite'' Higgs, the fermion couplings of $h$ $(\sigma)$ are not always a simple rescaling of the SM ones with $\cos\gamma$ $(\sin\gamma)$, but they are model-dependent;
\item[$\circ$] in a truly Twin Higgs model both $h$ and $\sigma$ are expected to have some invisible decay to particles of the SM-copy.
\end{itemize}
In the following we will assume a universal rescaling of the $h$ and $\sigma$ couplings. It is however important to emphasise the differences between the two scenarios, commenting on the impact of an invisible decay channel for $\sigma$ in the TH model.\footnote{In the case of TH, also $h$ can have an invisible decay width, which is mainly due to decay to mirror bottoms. It is almost always under control for allowed values of $f$. We will not comment on this any further.} In the assumption that the two copies of the SM have the same gauge and Yukawa couplings, the expression for the two vev's suggest that the particles of the invisible sector, which we indicate with a prime, have masses $m_{\rm SM'}\simeq m_{\rm SM}\times f/v$. Given the approximate $Z_2$ symmetry, the decay widths of $\sigma$ to invisible particles ({\it i.e.}\ in the SM-copy) are similar to those of the SM Higgs boson to the corresponding visible particles. More precisely,  the same predictions for the widths apply, upon substituting $v\mapsto f$. In particular, then, above the threshold $m_\sigma >2 m_{W'}\simeq 2 \times f/v \times m_W$ the invisible width can become dominant,
\begin{equation}
\Gamma_{\sigma \to \mathrm{inv.}}\simeq 3\,\Gamma_{\sigma \to Z'Z'}\simeq 3\,\Gamma_{\sigma \to ZZ} \simeq 3\,\Gamma_{\sigma \to hh},
\end{equation}
where the last two equalities hold because of the equivalence theorem, that now involves the 7 Goldstone bosons of our construction.
This means that, instead of the relation \eqref{eq:BRasymptotic}, one has
\begin{equation}
\mathrm{BR}_{\sigma \to \mathrm{inv.}}\simeq 3\,\mathrm{BR}_{\sigma \to ZZ}\simeq \frac{3}{7}.
\end{equation}
We will show the effect of the invisible branching ratios on direct searches, but we will also present results in the case $\mathrm{BR}_{\sigma \to \mathrm{inv.}}=0$, as a paradigmatic example of the composite Higgs picture.

The results are shown in \figuresnames s~\ref{fig:twin} and \ref{fig:chm}, where we present the case of a TH and a CH model respectively. The two scenarios differ mainly for the invisible branching ratio of the $\sigma$ particle. Moreover, as already discussed, in the CH case we are assuming an universal rescaling of all the SM Higgs couplings, although this a model-dependent issue in concrete models. 

In both cases, if the constraints are presented in a plane $(m_\sigma, f)$ we see that at each collider stage indirect and direct searches will provide complementary information. In particular direct ones will help probing a region of small masses (and reasonably small mixing) that is not accessible to the Higgs coupling fit, allowing for another approach to study the TH models rather than the simple analysis of the deviation in the Higgs couplings.  On the contrary the Higgs couplings put a robust bound for large values of $m_\sigma$, where the direct searches are not really effective{, and will be affected by the broadening of the resonance.
Notice that the BR$_{\sigma\to hh}$ is always fixed by \eqref{TH_Higgs_couplings}, differently from the generic and NMSSM cases, and implies that direct searches in TH and CHM are always dominated by the $VV$ channel.


\subsection{Electroweak precision tests}\label{sec:CHM_EWPT}

As already discussed, the IR-logarithms of \eqref{EWPT} give sizeable contributions to the $S$ and $T$ parameters \cite{Barbieri:2007bh}, which depend on the mixing $\sin^2\gamma$ and $m_\sigma$. In the case where the UV is mostly weakly coupled, hence for small values of the mass $m_\sigma \sim f$, EWPT are not an important constraint. 

On the contrary, in a strongly coupled scenario where $m_\sigma\sim 4\pi f$, EWPT are expected to be more relevant. In this limit $\sin^2\gamma\sim v^2/f^2$, while the logarithm in \eqref{EWPT} is maximal, $\log (4\pi f/ m_h)$, and EWPT put a constraint on $f$. This reproduces the usual model independent contribution of composite Higgs models, and it can be read off (\ref{EWPT}). In the absence of other contributions, the EWPT bound becomes comparable to the Higgs coupling constraints, also given the signs of the S and T contributions of (\ref{EWPT}).
In explicit composite models, tree level contributions from spin-1 resonances to the $S$ parameter make the bound even stronger \cite{Ciuchini:2013pca} (see \cite{Panico:2015jxa} for a review).
However, UV contributions can relax the bound from EWPT to a level comparable with the actual precision on Higgs couplings \cite{Grojean:2013qca,Azatov:2013ura,Thamm:2015zwa}, so we decide not to show this constraint, the result being more model-dependent.

It is also interesting to compare the future improvements in the Higgs couplings and EWPT (see also \cite{Barbieri:2013aza}). EWPT are expected to benefit from a period of $Z$-pole measurements at an $e^+e^-$ collider, such as FCC-ee and CEPC. Present studies \cite{Mishima,Fan:2014vta} suggest that the error on the $S$ and $T$ parameters (or the $\epsilon$-parameters) will shrink by an order of magnitude at a high-luminosity $Z$-pole facility. This improvement can be compared to the attainable precision on the $hZZ$ coupling through the absolute measure of $\sigma_{hZ}$ at a circular $e^+e^-$ collider, that is expected to strengthen the actual bounds by two order of magnitudes. They thus both reach a comparable level of precision on the scale $f$ \cite{Thamm:2015zwa}: for this reason, and given the model dependency of the EWPT, we do not show projections of such constraints.


\section{Conclusions}\label{sec:conclusions}

In this paper we have analysed the present and future constraints on models where a scalar singlet is added to the SM and is mixed with the Higgs boson. We focused on two main phenomenological aspects: direct searches for the new state at colliders, and indirect effects in the couplings of the 125 GeV Higgs due to its mixing with the singlet.

We put a particular attention to the prospects of direct searches at the various stages of the LHC -- the imminent 13-14 TeV run, and its possible luminosity and energy upgrades -- and at a future high-energy hadron collider. In order to do that we have used an approximate method, already presented in \cite{SalamWeiler,Thamm:2015zwa}, for extrapolating the existing constraints to higher energy and luminosity. The procedure is subject to several important assumptions and approximations: we have discussed them in detail and have performed various checks of our results, finding an overall agreement with the previous literature.

At the same time, we have considered the prospects for measuring deviations in the Higgs couplings, both at the LHC and at future linear and circular lepton colliders, trying to highlight the interplay between direct and indirect searches.

We first discussed the model-independent case of a generic singlet, with interactions not constrained by any symmetry.
The mixing of the SM Higgs boson with a scalar singlet allows an extremely simple description in terms of only two free parameters: the physical mass of the extra scalar and the mixing angle among the two states. These two quantities alone determine, in full generality, most of the phenomenology of the two particles, {\it i.e.}\ their production cross-sections and branching ratios into SM vectors and fermions. Interestingly enough, we have found that only one further parameter, the vev of the singlet $v_s$, is needed to describe the remaining properties of the Higgs system, namely the couplings among the scalars. Furthermore, the branching ratio of the singlet to a pair of Higgs bosons is fixed by the equivalence theorem at high masses, thus reducing effectively the total number of parameters to two in this limit.

The relevant direct searches are always the ones for resonances in the $W^+W^-$, $ZZ$ and $hh$ channels. Their relative importance depends on the size of ${\rm BR}_{\phi\to hh}$, although the $VV$ channel is typically the most sensitive.
When compared with Higgs coupling measurements, direct searches will always provide a useful complementary tool; we find that, in general, they dominate for lower masses of the scalar singlet. An interesting point that emerged is the importance of the Higgs self-coupling: there are regions of the parameter space where large deviations from the SM could show up, even without observable effects in the Higgs signal strengths. This not only motivates an effort to optimise the reach at future machines, but also indicates that deviations could hide in the present and soon-to-come data, which we thus encourage to analyse in this respect.

We have then specialised our analysis to a couple of interesting models that are expected to provide an extra scalar singlet at sufficiently low scales: the NMSSM and Twin Higgs and Composite Higgs models. These are at present also the best candidates for physics beyond the SM that are able to accommodate an (almost) natural electroweak scale with the various constraints from the LHC.
Their description deviates from the previous generic case since there are approximate or softly broken symmetries that constrain the interactions in the scalar potential.
We then have found different limits with respect to the above case.

\begin{itemize}
\item[\textsc{1)}] The NMSSM with a light CP-even singlet. In this case direct searches provide an useful lamppost to find signs of new physics: we find it interesting that they are more powerful than the fit to the Higgs couplings. We considered the two cases of a perturbative and a strong coupling between the Higgs fields, the main difference among them is that the first case is more difficult to probe.
Like in the case of a generic singlet, there is room for a sizeable modification of the trilinear Higgs coupling.
\item[\textsc{2)}] Twin \& Composite Higgs with a light ``radial mode''. Although having a different physical interpretation, these two cases both provide a scenario where the Higgs boson is a pseudo-Goldstone boson of a spontaneously broken symmetry, and can be described in complete similarity.
The real difference here is that the mixing angle does not go to zero with the mass of the singlet, providing a robust bound from the Higgs couplings on a large range of the parameter space. We have however found that if the mass of the radial mode is sufficiently small -- a motivated assumption in a weakly coupled Twin Higgs scenario -- then direct searches are stronger than the Higgs couplings.
\end{itemize}

It is clear that the presence of an extra singlet scalar close to the weak scale is an interesting case to be studied. Despite its simplicity, it can offer diverse phenomenological scenarios, with interesting prospects for any future experiment both at lepton and hadron colliders. On the other hand, even without looking too far ahead, we have shown that already the second run of the LHC can efficiently probe this scenario, and we do hope that a great effort will be put by the experimental collaborations to close in on this picture.


\subsubsection*{Acknowledgements}
We thank Andrea Thamm and Riccardo Torre for insightful discussions about the extrapolation procedure. We also acknowledge useful conversations with Riccardo Barbieri, Matthew Low, Caterina Vernieri and Lian-Tao Wang. A special thank goes to Andrea Romanino and Andrea Wulzer for the possibility to join the INFN initiative \textit{``What Next''} which has triggered part of this work.
FS acknowledges the hospitality of the Institut d'Astrophysique de Paris ({\sc Iap}).
DB is financed by the European Research Council in the context of the {\sc Erc} Advanced Grant project `{\sc Flavour}' (267104).
FS is supported by the European Research Council ({\sc Erc}) under the EU Seventh Framework Programme (FP7 2007-2013)/{\sc Erc} Starting Grant (agreement n.\ 278234 --- `{\sc NewDark}' project). The work of AT was supported by an Oehme Fellowship.



\appendix

\section{More on the extrapolation method}\label{sec:appendix}

In order to discuss the issues related to our extrapolation procedure, we would like to provide here more details about some of the steps already presented in \sectionname~\ref{sec:method}.

The estimation of the background relies on \eqref{background}, where the main quantities are the SM cross-section $\hat{\sigma}_{ij}$ and the parton luminosities $\mathrm{d} \mathcal{L}_{ij}/ \mathrm{d} \hat{s}$. The latter can be computed as \cite{Ellis:1991qj}
\begin{equation}\label{PDF}
\frac{\d\L_{ij}}{\d\hat s}(\hat s,s)\equiv \frac{1}{1+\delta_{ij}} \frac{1}{s} \int_{\hat{s}/s}^1 \frac{\d x}{x} \Big[ f_i \big(x, \hat{s}\big)f_j \big(\frac{\hat{s}}{x s}, \hat{s}\big) + (i \leftrightarrow j) \Big], 
\end{equation}
where $f_i(x,Q^2)$ is the parton distribution function (PDF) of the parton $i$, and the factorisation scale $Q^2$ has always been fixed to $Q^2=\hat{s}$. For the background scaling with $\hat{s}$ to be driven by the parton luminosities only, one needs the cross-section to behave like
\begin{equation}\label{approx-cqq}
\hat{s}\,\sigma_{ij} \cong c_{ij}
\end{equation}
above a certain $\hat{s}$ value. Threshold effects due to SM particles decouple like $m_{\rm SM}^2/\hat{s}$, where $m_{\rm SM}$ is the mass of the particle relevant for the background, {\it e.g.}\ twice the top mass for $pp \to 4b$. When going above those thresholds, one could still be worried by the presence of logarithms, that are {\it e.g.}\ present whenever there is a contribution from a $t$-channel process. This is for example the case for $q\bar q\to ZZ,W^+W^-$, but also $gg\to q\bar q$, which are all relevant backgrounds for the processes that we are interested in. Such logarithmic contributions are expected to be subdominant, because the experimental cuts used in the above searches tend to exclude the singular kinematic configurations which are responsible for them. Nonetheless, we have explicitly checked that the logarithms are not important at current energies, by verifying that the excluded cross-sections scale with the parton luminosities to an extremely good approximation (see \figuresnames~\ref{scaling}).\footnote{In fact if the approximation of \eqref{approx-cqq} would not be verified, then one would see deviations in the above scaling due to the neglected logarithms.}
Finally, we verified that one can safely neglect the same logarithms also at higher energies, at least in the range relevant in this paper, by virtue of their weak dependence on $\hat s$.

We now discuss the parton luminosities used for the rescaling of the backgrounds relevant for our direct searches. The background for the process $\phi \to ZZ$, which leads the $VV$ combination \cite{Khachatryan:2015cwa}, is dominated at the partonic level by $q\bar{q} \to ZZ$ \cite{Chatrchyan:2013mxa,Khachatryan:2015cwa}. 
Matching \eqref{approx-cqq} with the tree-level scattering cross-section $q\bar{q} \to ZZ$, one obtains that the coefficients for the up and down-type quarks satisfy the relation $ c_{u\bar u}/c_{d\bar d} \simeq (g_{Lu}^4+g_{Ru}^4)/(g_{Ld}^4+g_{Rd}^4)\simeq 0.5$, where $g_{Lq}$ and $g_{Rq}$ are the $Z$ couplings to the left- and right-handed quark currents. A further simplification can be achieved observing that, on the relevant mass scales at each collider, the parton luminosities of the up and down quark are in an approximate constant ratio. This means that the background $q\bar{q}\to ZZ$ can be rescaled to an extremely good approximation by considering either $\d\L_{d\bar d}/\d\hat{s}$ or $\d\L_{u\bar u}/\d\hat{s}$, given that in \eqref{final_bkg} any constant factor cancels out. 
Concerning now the other relevant search, the background to the process $\phi \to hh(4b)$ is consituted by $t\bar t$ and multijet events \cite{Khachatryan:2015yea}, and pinning down its precise partonic composition is an involved task. However we expect an important contribution to come from the $gg$ initial state, and we have chosen to perform the extrapolation using the $gg$ parton luminosity. We have nevertheless checked that the impact of choosing another parton luminosity on the exclusions presented in this paper is mild.

In \figuresnames~\ref{scaling} we show a comparison between the 95\% C.L. expected exclusions on the cross-section times branching ratio, from LHC8, and the square-root of the (arbitrarily normalised) parton luminosities used for the corresponding extrapolations. The left panel shows the exclusion in the $\phi\to VV$ channel together with the $\L_{d\bar d}$ parton luminosity, while the right panel shows the exclusion in the $\phi\to hh(4b)$ channel compared with the $\L_{gg}$ luminosity. One sees that, for masses $m_\phi$ above the relevant SM thresholds, the agreement between the two curves is extremely good. This confirms at the same time that the exclusion on the cross-section is determined mainly by the number of background events, and that the latter is proportional to the parton luminosities times a constant factor, thus substantiating equation \eqref{approx-cqq}.
In other words, \figuresnames~\ref{scaling} provides a check that the approximation \eqref{approx-cqq} holds at the LHC8, and gives the further information of the mass from which it is valid.

\begin{figure}[t]
\centering%
\includegraphics[width=0.49\textwidth]{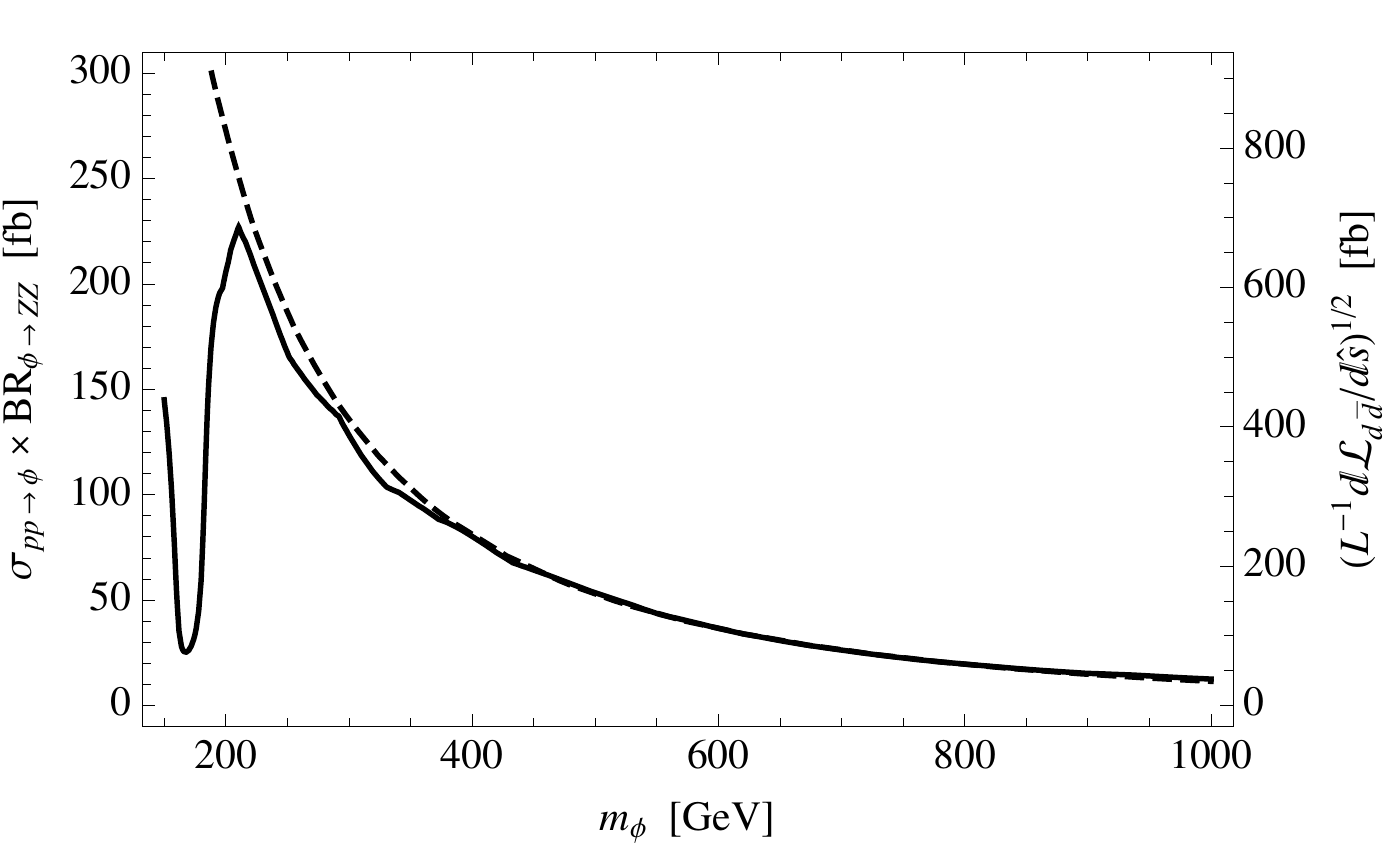}\hfill%
\includegraphics[width=0.49\textwidth]{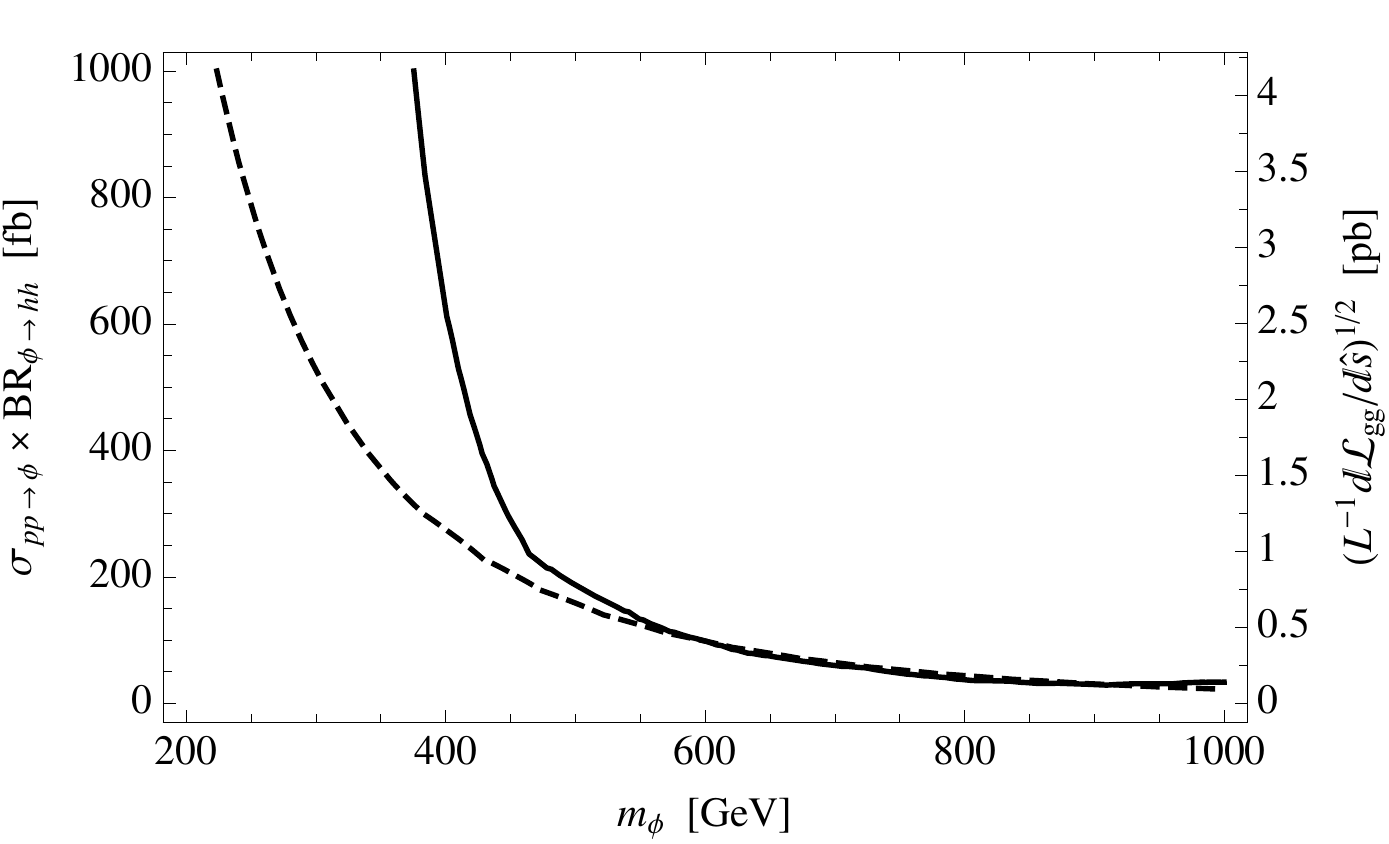}
\caption{Left: comparison between the LHC8 expected exclusion on $\sigma_{pp\to\phi}\times{\rm BR}_{\phi\to VV}$ (solid) and $\sqrt{L^{-1}\d\L_{d\bar d}/\d \hat s}$ (dashed). Right: the same as before, but with ${\rm BR}_{\phi\to hh}$ and $\L_{gg}$.\label{scaling}}
\end{figure}

\pagestyle{plain}
\bibliographystyle{jhep}
\small
\bibliography{ref}

\end{document}